\documentclass[aps,pra,twocolumn,superscriptaddress,showpacs,floatfix]{revtex4-2}
\usepackage{amsmath,amssymb,bm,graphicx,color}
\usepackage[colorlinks=true, linkcolor=blue, citecolor=blue, urlcolor=blue]{hyperref}
\usepackage{braket}
\usepackage{physics}
\usepackage{siunitx}
\usepackage{float}
\usepackage{subcaption}

\begin{document}

\title{Reliability Dynamics in a Two-Site Dissipative Quantum Spin Chain}

\author{Bowen Sun}
\affiliation{Institute of Physics, Beijing National Laboratory for Condensed Matter Physics,\\
Chinese Academy of Sciences, Beijing 100190, China}
\affiliation{School of Physical Sciences, University of Chinese Academy of Sciences, Beijing 100049, China}

\author{D. L. Zhou} 
\email[]{zhoudl72@iphy.ac.cn}
\affiliation{Institute of Physics, Beijing National Laboratory for Condensed Matter Physics,\\
Chinese Academy of Sciences, Beijing 100190, China}
\affiliation{School of Physical Sciences, University of Chinese Academy of Sciences, Beijing 100049, China}

\date{\today}

%===============================================================================
% Abstract
%===============================================================================
\begin{abstract}
As a key index for applications of a device, the device's reliability is its ability to survive (function normally over time) under the influence of some environment.
In this paper we present a quantum energy-storing device model with a quantum spin chain, whose environment influence is described by the Lindblad master equation.
Here the device survives if the spin system stays in the state with nonzero excitations, otherwise it fails. 
Because the Lindblad dynamics enforces one-way energy decay and strict irreversibility of the state of failure, we are capable of investigating the reliability of the quantum device directly using classical reliability theory. 
Focusing on the minimal nontrivial case---a two-site spin-$1/2$ chain, we derive closed-form expressions for the reliability and the hazard.
The dynamics exhibit an overdamped--underdamped crossover controlled by the competition between coherent exchange and dissipation inhomogeneity.
The exact analytical formulas are in excellent agreement with numerical simulations. 
More importantly, we establish an experimentally accessible protocol for assessing reliability based on first-passage time statistics.

\end{abstract}

\maketitle

%===============================================================================
% Section 1:Introduction
%===============================================================================
\section{Introduction}

Reliability characterizes the probability that a system continues operating without failure, serving as a cornerstone concept in classical engineering and systems theory~\cite{rausand2020system, jaynes1957informationI, jaynes1957informationII, zio2009reliability, rocchi2017reliability, bain2017statistical, du2020maximum, pham2023springer}.
In essence, it quantifies the ability of a device to survive and maintain its functionality under environmental influence over time.

The rapid development of quantum technologies-including quantum computers~\cite{lloyd1996universal, preskill2018quantum, altman2021quantum}, quantum communication networks~\cite{kimble2008quantum, wehner2018quantum, azuma2023quantum}, and quantum sensors~\cite{degen2017quantum}-has created a growing need to understand the stability and long-term performance of quantum devices.
Extending reliability concepts to such systems is therefore a natural step toward characterizing their robustness~\cite{LIN2020126207, cui2023quantum, weinbrenner2024aging, cui2025quantum, liu2025optimal, li2025constructing, yi2025capacity}.
However, quantum coherence, superposition, and stochastic measurement outcomes introduce conceptual subtleties absent in classical reliability theory~\cite{nielsen2010quantum}.

A previous work introduced trajectory-based reliability measures by defining survival and failure subspaces through projection operators and evaluating their expectation values along quantum trajectories~\cite{cui2023quantum}.
Although this framework provides a natural quantum generalization of classical reliability, it reveals a key difficulty:
because quantum trajectories under continuous monitoring remain reversible, a system that has entered the failure subspace may stochastically return to the survival one.
Such post-failure recovery has no classical analogue without explicit repair and therefore obscures the operational meaning of reliability as an irreversible quantity.

In this work we enforce irreversibility directly at the dynamical level by considering Lindblad evolution with amplitude damping~\cite{lindblad1976generators, breuer2002theory, PRXQuantum.5.020202}.
A natural physical realization of this framework is a quantum energy-storing device, where stored excitations represent the operational capacity of the system and failure corresponds to the complete loss of excitations.
Under amplitude-damping dynamics, excitations can only decay and never spontaneously reappear, so that the fully decayed ground state becomes an absorbing failure state.
This property allows the reliability of the quantum device to be analyzed within the standard framework of classical reliability theory.
We study a dissipative quantum spin chain where the total excitation number $N_\uparrow$ quantifies the device's working capacity:
states with $N_\uparrow>0$ represent operational configurations, while the ground state with $N_\uparrow=0$ corresponds to device failure.

To obtain a clear analytical benchmark, we focus on the minimal nontrivial case: a two-site spin-$1/2$ chain with nonuniform dissipation.
Despite its simplicity, this model already captures the essential competition between coherent exchange and dissipation inhomogeneity.
It thus serves as a useful setting for establishing the analytical theory, benchmarking it against Lindblad numerics, and developing experimentally accessible reliability-assessment protocols based on first-passage statistics.

The remainder of this paper is organized as follows.
Section~II introduces the basic concepts in reliability framework.
Section~III presents the dissipative two-site spin model and formulates the Lindblad master equation.
Section~IV derives the closed equations of motion governing the reliability dynamics.
Section~V provides closed-form expressions for the reliability function $R(t)$ and the hazard $h(t)$, and analyzes the overdamped--underdamped crossover.
Section~VI validates the analytical results with Lindblad numerics.
Section~VII develops experimentally accessible assessment protocols based on first-passage statistics and analyzes the associated estimator fluctuations.
Section~VIII concludes and outlines possible directions for future research.

%===============================================================================
% Section 2:Reliability Framework
%===============================================================================
\section{Reliability Framework}

In classical reliability theory, the reliability function $R(t)$ characterizes the probability that a system remains operational up to time $t$, i.e.,
\begin{equation}
R(t) = P(T > t),
\end{equation}
where $T$ is a random variable describing the system's failure time. The complementary probability,
\begin{equation}
F(t) = P(T \le t),
\end{equation}
represents the cumulative distribution of failure events, such that
\begin{equation}
R(t) = 1 - F(t).
\end{equation}

The failure rate or hazard function $h(t)$ measures the instantaneous probability per unit time that the system fails at time $t$, given that it has survived up to this point:
\begin{equation}
h(t) = -\frac{d}{dt} \ln R(t)
      = \frac{f(t)}{R(t)},
\end{equation}
where $f(t) = \frac{dF(t)}{dt} = -\frac{dR(t)}{dt}$ denotes the probability density function of failure times.

The reliability function $R(t)$ and the hazard function $h(t)$ are the two core quantities in reliability theory: $R(t)$ characterizes the survival probability, while $h(t)$ describes the instantaneous failure tendency, and knowledge of either one determines the other.

An exponential reliability function $R(t) = e^{-\lambda t}$ corresponds to a constant hazard $h(t) = \lambda$, describing a memoryless process typical of Poissonian failure statistics~\cite{rausand2020system}. 
In contrast, nonexponential forms of $R(t)$ or time-dependent $h(t)$ indicate more complex behaviors such as aging, fatigue, or heterogeneous dissipation.

For the quantum energy-storing device considered here, $R(t)$ is identified with the probability that the system has not reached the absorbing ground state.

%===============================================================================
% Section 3:Two-Site Dissipative Spin Model
%===============================================================================
\section{Two-Site Dissipative Spin Model}
%===============================================================================
% Subsection 3.1:Hamiltonian
%===============================================================================
\subsection{Hamiltonian}

We consider a spin-$1/2$ chain with coherent nearest-neighbor exchange, described by
\begin{equation}
H
=
\sum_{i=1}^{N} \epsilon_i \sigma_z^{(i)}
+
J \sum_{i=1}^{N-1}
\left(
\sigma_+^{(i)}\sigma_-^{(i+1)}
+
\sigma_+^{(i+1)}\sigma_-^{(i)}
\right),
\label{eq:H_general}
\end{equation}
where $\epsilon_i$ denotes the on-site energy of site $i$, and $J$ is the coherent exchange coupling strength.

Define the total magnetization and excitation-number operators
\begin{equation}
S_z = \sum_{i=1}^{N} \sigma_z^{(i)},
\qquad
N_\uparrow = \sum_{i=1}^{N} \sigma_+^{(i)}\sigma_-^{(i)} .
\end{equation}
Using the commutation relations $[\sigma_z^{(i)},\sigma_\pm^{(j)}]=\pm 2\delta_{ij}\sigma_\pm^{(i)}$, one verifies
\begin{equation}
[H,S_z]=0,
\qquad
[H,N_\uparrow]=0.
\label{eq:coherent_commute}
\end{equation}
Hence the coherent dynamics preserves the total excitation number: the exchange term only transfers excitations between neighboring sites without creating or annihilating them.

In this work we focus on the minimal nontrivial case $N=2$, for which
\begin{equation}
H
= \epsilon_1 \sigma_z^{(1)} + \epsilon_2 \sigma_z^{(2)}
+ J\left(\sigma_+^{(1)}\sigma_-^{(2)} + \sigma_+^{(2)}\sigma_-^{(1)}\right).
\label{eq:H}
\end{equation}
The Hilbert space then decomposes into invariant excitation-number sectors,
\begin{equation}
\mathcal{H}=\bigoplus_{k=0}^{2}\mathcal{H}_k,
\qquad
\mathcal{H}_k=\mathrm{span}\Big\{\,|\phi\rangle:\ N_\uparrow|\phi\rangle=k|\phi\rangle\,\Big\},
\end{equation}
with dimensions $\dim\mathcal{H}_0=1$, $\dim\mathcal{H}_1=2$, and $\dim\mathcal{H}_2=1$.
Explicitly,
\begin{equation}
\begin{aligned}
\mathcal{H}_0 &= \mathrm{span}\{\ket{00}\},\\
\mathcal{H}_1 &= \mathrm{span}\{\ket{10},\ket{01}\},\\
\mathcal{H}_2 &= \mathrm{span}\{\ket{11}\}.
\end{aligned}
\end{equation}
Here $\ket{10}$ ($\ket{01}$) denotes a single excitation localized on site $1$ ($2$).
In this basis, the Hamiltonian is block diagonal,
\begin{equation}
H \;=\; H_0 \oplus H_1 \oplus H_2,
\end{equation}
where $H_0$ and $H_2$ are one-dimensional, while the nontrivial coherent mixing occurs entirely within the single-excitation sector $\mathcal{H}_1$ through the exchange coupling $J$.

A finite detuning $\Delta\epsilon=\epsilon_1-\epsilon_2$ introduces an additional relative phase rotation between the single-excitation basis states $\ket{10}$ and $\ket{01}$.
For clarity and to keep the analytical expressions compact, we focus on the resonant case $\Delta\epsilon=0$ in what follows; including $\Delta\epsilon\neq0$ is straightforward and does not change the dissipation--exchange competition that underlies the crossover behavior of interest.

%===============================================================================
% Subsection 3.2:Dissipation and Lindblad equation
%===============================================================================
\subsection{Dissipation and Lindblad equation}

Each spin undergoes local amplitude damping with rates $\gamma_1,\gamma_2\,(\gamma_1>0,\gamma_2>0)$,
\begin{equation}
L_1=\sqrt{\gamma_1}\,\sigma_-^{(1)},\qquad
L_2=\sqrt{\gamma_2}\,\sigma_-^{(2)}.
\end{equation}
The system dynamics is governed by the Lindblad master equation~\cite{lindblad1976generators}
\begin{equation}
\dot{\rho}
=-i[H,\rho]
+\sum_{i=1}^2\left(L_i\rho L_i^\dagger -\frac12\{L_i^\dagger L_i,\rho\}\right).
\label{eq:ME}
\end{equation}

Here we adopt an effective local-dissipation description.
That is, we retain only the on-site amplitude-damping channels and neglect bath-induced cross-dissipative terms that may arise in more microscopic treatments with a common reservoir or correlated baths~\cite{wang2015dissipation}.
Accordingly, the present master equation should be understood as a simplified local Lindblad model, appropriate when spatial interference effects of the environment are ignored and only the effective local decay rates are kept.

We define the failure state as the global ground state $\ket{00}$, which is absorbing under Lindblad dynamics.
The failure probability is $F(t)=\bra{00}\rho(t)\ket{00}$ and hence the reliability function is
\begin{equation}
R(t)=1-\bra{00}\rho(t)\ket{00}.
\label{eq:Rdef}
\end{equation}

%===============================================================================
% Section 4:Equations of Motion
%===============================================================================
\section{Equations of Motion}
\label{sec:eom}

It is convenient to view the Lindblad master equation~\eqref{eq:ME} as a linear equation generated by a Liouville superoperator~\cite{PRXQuantum.5.020202},
\begin{equation}
\dot{\rho}(t)=\mathcal{L}\big[\rho(t)\big],
\label{eq:Liouville_form_novec}
\end{equation}
where $\mathcal{L}$ is a linear map acting on the density operator.
Once a basis is chosen, Eq.~\eqref{eq:Liouville_form_novec} is equivalent to a self-contained system of linear ordinary differential equations for the density-matrix elements.

To express the master equation in component form, we use the ordered basis
\begin{align}
\ket{0}=\ket{00},\quad
\ket{1}=\ket{10},\nonumber\\
\ket{2}=\ket{01},\quad
\ket{3}=\ket{11},
\end{align}
and denote density-matrix elements by $\rho_{mn}=\bra{m}\rho\ket{n}$.
The structure of the component equations is strongly constrained by the dissipator.
For local amplitude damping $L_i=\sqrt{\gamma_i}\,\sigma_-^{(i)}$,
\begin{equation}
L_i^\dagger L_i=\gamma_i\,\sigma_+^{(i)}\sigma_-^{(i)}\equiv \gamma_i n_i,
\end{equation}
where $\sigma_-^{(i)}$ lowers the excitation number by one while $n_i=\sigma_+^{(i)}\sigma_-^{(i)}$ is the local excitation-number operator and therefore does not change the excitation number.
Consequently, the jump term $L_i\rho L_i^\dagger$ can only transfer population from higher-excitation states to lower-excitation states (e.g., $\ket{11}\to\ket{10},\ket{01}$ and $\ket{10},\ket{01}\to\ket{00}$), whereas the anticommutator term $-\tfrac12\{L_i^\dagger L_i,\rho\}$ generates decay of population and coherence within a given excitation sector.
This one-way decay structure yields a hierarchical set of component equations, where the higher-excitation dynamics act as sources for lower-excitation sectors but not vice versa.

Evaluating Eq.~\eqref{eq:ME} in the chosen basis, we obtain the equations of motion for the populations, which are coupled to the coherences:
\begin{align}
\dot{\rho}_{33} &= -(\gamma_1+\gamma_2)\rho_{33},
\label{eq:r33}\\
\dot{\rho}_{11} &= -\gamma_1\rho_{11}
- iJ(\rho_{21}-\rho_{12})
+ \gamma_2\rho_{33},
\label{eq:r11}\\
\dot{\rho}_{22} &= -\gamma_2\rho_{22}
- iJ(\rho_{12}-\rho_{21})
+ \gamma_1\rho_{33},
\label{eq:r22}\\
\dot{\rho}_{00} &= \gamma_1\rho_{11}+\gamma_2\rho_{22}.
\label{eq:r00}
\end{align}
Equation~\eqref{eq:r00} shows explicitly that $\ket{00}$ is an absorbing state:
population only enters $\rho_{00}$ and never leaves. Moreover, the above equations are consistent with the normalization condition $\mathrm{Tr}\rho(t)=1$, namely,
\begin{equation}
\rho_{00}(t)+\rho_{11}(t)+\rho_{22}(t)+\rho_{33}(t)=1.
\label{eq:trace}
\end{equation}

The exchange interaction in the Hamiltonian coherently couples the two single-excitation basis states $\ket{10}$ and $\ket{01}$. 
As a result, the populations $\rho_{11}$ and $\rho_{22}$ are coupled to the single-excitation coherence through the combination $i(\rho_{12}-\rho_{21})$.
For later convenience, we introduce the real quantity
\begin{equation}
\rho_m \equiv i(\rho_{12}-\rho_{21}) = 2\,\Im(\rho_{12}).
\end{equation}
In terms of $\rho_m$, Eqs.~\eqref{eq:r11}--\eqref{eq:r22} become
\begin{align}
\dot{\rho}_{11} &= -\gamma_1\rho_{11} + \gamma_2\rho_{33} + J\rho_m,
\label{eq:r11_Pm}\\
\dot{\rho}_{22} &= -\gamma_2\rho_{22} + \gamma_1\rho_{33} - J\rho_m.
\label{eq:r22_Pm}
\end{align}

The coherence equation obtained from Eq.~\eqref{eq:ME} reads
\begin{equation}
\dot{\rho}_{12}
=
-\frac{\gamma_1+\gamma_2}{2}\rho_{12}
-iJ(\rho_{22}-\rho_{11}).
\label{eq:r12}
\end{equation}
Combining this equation with its Hermitian conjugate yields
\begin{equation}
\dot{\rho}_m
=
-\frac{\gamma_1+\gamma_2}{2}\rho_m
+2J(\rho_{22}-\rho_{11}).
\label{eq:Pm_general}
\end{equation}
Introducing $\bar{\gamma}\equiv(\gamma_1+\gamma_2)/2$, this becomes
\begin{equation}
\dot{\rho}_m
=
-\bar{\gamma}\rho_m
+2J(\rho_{22}-\rho_{11}).
\label{eq:Pm_resonant}
\end{equation}

Together with Eq.~\eqref{eq:r33} (while $\rho_{00}$ follows from normalization), the dynamics close on the four-component vector
\begin{equation}
\bm{x}(t)\equiv\big(\rho_{11}(t),\,\rho_{22}(t),\,\rho_{33}(t),\,\rho_m(t)\big)^T,
\end{equation}
which satisfies the linear system
\begin{equation}
\dot{\bm{x}}(t)=A\,\bm{x}(t),
\label{eq:x_Ax}
\end{equation}
with
\begin{equation}
A=
\begin{pmatrix}
-\gamma_1 & 0 & \gamma_2 & J\\
0 & -\gamma_2 & \gamma_1 & -J\\
0 & 0 & -2\bar{\gamma} & 0\\
-2J & 2J & 0 & -\bar{\gamma}
\end{pmatrix}.
\label{eq:A_matrix}
\end{equation}
This closed linear system forms the starting point for the exact analytical expressions of the reliability $R(t)$ and the hazard $h(t)$.

%===============================================================================
% Section 5:Closed-form Expressions for R(t) and h(t)
%===============================================================================
\section{Closed-form Expressions for \texorpdfstring{$R(t)$ and $h(t)$}{R(t) and h(t)}}
\label{sec:closed_form}

We now derive closed-form expressions for the reliability $R(t)$ and the hazard $h(t)$ directly from the component equations of motion obtained in Sec.~\ref{sec:eom}. 

%===============================================================================
% Subsection 5.1:General Formula
%===============================================================================
\subsection{General Formula}

The reliability is
\begin{equation}
R(t)=1-\rho_{00}(t)=\rho_{11}(t)+\rho_{22}(t)+\rho_{33}(t),
\label{eq:R_def_here}
\end{equation}
where the second equality follows from normalization.
Combining Eqs.~\eqref{eq:r33}, \eqref{eq:r11}, and \eqref{eq:r22}, we obtain
\begin{equation}
\dot{R}(t)= -\gamma_1\rho_{11}(t)-\gamma_2\rho_{22}(t).
\label{eq:Rdot_key}
\end{equation}
Equivalently, Eq.~\eqref{eq:r00} gives $\dot{\rho}_{00}(t)=-\dot{R}(t)$.

Hence,
\begin{equation}
h(t)= -\frac{\dot R(t)}{R(t)}
=
\frac{\gamma_1\rho_{11}(t)+\gamma_2\rho_{22}(t)}
{\rho_{11}(t)+\rho_{22}(t)+\rho_{33}(t)}.
\label{eq:hazard_general}
\end{equation}

%===============================================================================
% Subsection 5.2:Spectrum of \texorpdfstring{$A$}{A} and general solution structure
%===============================================================================
\subsection{Spectrum of \texorpdfstring{$A$}{A} and mode expansion}
\label{subsec:spectrum}

Recall that $\bm x(t)=(\rho_{11},\rho_{22},\rho_{33},\rho_m)^{\mathsf T}$ satisfies the linear system $\dot{\bm x}(t)=A\bm x(t)$ with $A$ given in Eq.~\eqref{eq:A_matrix}.
Let $\bm v_k$ denote the eigenvector of $A$ associated with eigenvalue $\lambda_k$,
\begin{align}
A\bm v_k=\lambda_k \bm v_k,
\end{align}
defined up to an overall nonzero scalar factor.

The eigenvalues of $A$ are
\begin{align}
\lambda_1&=-\bar{\gamma},\nonumber\\
\lambda_2&=-2\bar{\gamma},\nonumber\\
\lambda_3&=-\bar{\gamma}-\frac{\Lambda}{2},\nonumber\\
\lambda_4&=-\bar{\gamma}+\frac{\Lambda}{2},
\label{eq:eigs_A}
\end{align}
where $\Lambda\equiv\sqrt{(\Delta\gamma)^2-16J^2}$ and $\Delta\gamma\equiv\gamma_1-\gamma_2$.

A convenient choice of the corresponding eigenvectors is
\begin{align}
\bm v_1 &= \left(\frac{2J}{\Delta\gamma},\,\frac{2J}{\Delta\gamma},\,0,\,1\right)^{\mathsf{T}},\nonumber\\
\bm v_2 &= \left(-1,\,-1,\,1,\,0\right)^{\mathsf{T}},\nonumber\\
\bm v_3 &= \left(
\frac{\Lambda}{4J}+\frac{2J}{\Delta\gamma+\Lambda},\,
\frac{2J}{\Delta\gamma+\Lambda},\,
0,\,1\right)^{\mathsf{T}},\nonumber\\
\bm v_4 &= \left(
-\frac{\Lambda}{4J}+\frac{2J}{\Delta\gamma-\Lambda},\,
\frac{2J}{\Delta\gamma-\Lambda},\,
0,\,1\right)^{\mathsf{T}}.
\label{eq:evecs_A}
\end{align}

The general solution therefore admits the eigenmode expansion
\begin{align}
\bm x(t)=\sum_{k=1}^{4} C_k\,e^{\lambda_k t}\,\bm v_k,
\label{eq:x_general_modes}
\end{align}
where the coefficients $\{C_k\}$ are determined by the initial condition $\bm x(0)$ through
\begin{align}
\bm x(0)=\sum_{k=1}^{4} C_k\,\bm v_k.
\label{eq:C_from_x0}
\end{align}
Equivalently, the solution can be written in matrix form as $\bm x(t)=e^{At}\bm x(0)$.
%===============================================================================
% Subsection 5.3:Explicit solution for the |11> initial state
%===============================================================================
\subsection{Explicit solution for the \texorpdfstring{$\ket{11}$}{11} initial state}
\label{subsec:init11}

We now determine the mode amplitudes $\{C_k\}$ in Eq.~\eqref{eq:x_general_modes}
for the experimentally relevant initial state $\rho(0)=\ket{11}\!\bra{11}$.
In the reduced variables this corresponds to
\begin{align}
\bm x(0)=(\rho_{11}(0),\rho_{22}(0),\rho_{33}(0),\rho_m(0))^{\mathsf T}
=(0,0,1,0)^{\mathsf T}.
\label{eq:x0_ket11}
\end{align}
Substituting Eq.~\eqref{eq:x0_ket11} into the matching condition
\eqref{eq:C_from_x0} yields
\begin{align}
(0,0,1,0)^{\mathsf T}
=\sum_{k=1}^{4} C_k\,\bm v_k .
\label{eq:lin_C_ket11}
\end{align}

Using the eigenvectors \eqref{eq:evecs_A}, this system immediately gives
\begin{align}
C_2=1, \qquad
C_3=C_4, \qquad
C_1=-2C_3 .
\end{align}
Substituting these relations back into Eq.~\eqref{eq:lin_C_ket11} determines the coefficients
\begin{align}
C_1=-\frac{8J\Delta\gamma}{\Lambda^2},\qquad
C_2=1,\qquad
C_3=C_4=\frac{4J\Delta\gamma}{\Lambda^2},
\label{eq:Ck_ket11_final}
\end{align}
where $(\Delta\gamma)^2-\Lambda^2=16J^2$ has been used.

Substituting Eq.~\eqref{eq:Ck_ket11_final} into the mode expansion
\eqref{eq:x_general_modes} yields the explicit time dependence of the components
\begin{align}
\rho_{11}(t)
&=-\frac{16J^2}{\Lambda^2}e^{-\bar\gamma t}
-e^{-2\bar\gamma t}\nonumber\\
&\quad
+\frac{\Delta\gamma(\Delta\gamma+\Lambda)}{2\Lambda^2}
e^{-(\bar\gamma+\Lambda/2)t}\nonumber\\
&\quad
+\frac{\Delta\gamma(\Delta\gamma-\Lambda)}{2\Lambda^2}
e^{-(\bar\gamma-\Lambda/2)t},
\label{eq:rho11_from_modes}
\\
\rho_{22}(t)
&=-\frac{16J^2}{\Lambda^2}e^{-\bar\gamma t}
-e^{-2\bar\gamma t}\nonumber\\
&\quad
+\frac{\Delta\gamma(\Delta\gamma-\Lambda)}{2\Lambda^2}
e^{-(\bar\gamma+\Lambda/2)t}\nonumber\\
&\quad
+\frac{\Delta\gamma(\Delta\gamma+\Lambda)}{2\Lambda^2}
e^{-(\bar\gamma-\Lambda/2)t},
\label{eq:rho22_from_modes}
\\
\rho_{33}(t)
&=e^{-2\bar\gamma t}.
\label{eq:rho33_from_modes}
\end{align}

Using $R(t)=\rho_{11}(t)+\rho_{22}(t)+\rho_{33}(t)$ we obtain the reliability in closed form as
\begin{align}
R(t)
&=
-\frac{32J^2}{\Lambda^2}e^{-\bar\gamma t}
-e^{-2\bar\gamma t}\nonumber\\
&\quad
+\frac{(\Delta\gamma)^2}{\Lambda^2}
e^{-(\bar\gamma-\Lambda/2)t}\nonumber\\
&\quad
+\frac{(\Delta\gamma)^2}{\Lambda^2}
e^{-(\bar\gamma+\Lambda/2)t}.
\label{eq:R_ket11_closed}
\end{align}

Differentiating Eq.~\eqref{eq:R_ket11_closed} yields
\begin{align}
\dot R(t)
&=
\frac{32J^2}{\Lambda^2}\bar\gamma e^{-\bar\gamma t}
+2\bar\gamma e^{-2\bar\gamma t}\nonumber\\
&\quad
-\frac{(\Delta\gamma)^2}{\Lambda^2}
\left(\bar\gamma-\frac{\Lambda}{2}\right)
e^{-(\bar\gamma-\Lambda/2)t}\nonumber\\
&\quad
-\frac{(\Delta\gamma)^2}{\Lambda^2}
\left(\bar\gamma+\frac{\Lambda}{2}\right)
e^{-(\bar\gamma+\Lambda/2)t}.
\label{eq:Rdot_ket11_closed}
\end{align}

As a consistency check, $\dot R(t)$ also satisfies the exact identity
\eqref{eq:Rdot_key}.
The hazard then follows directly from Eq.~\eqref{eq:hazard_general}.
A key finding revealed by the closed-form expressions is that the competition between dissipation inhomogeneity and coherent exchange is encoded in $\Lambda=\sqrt{(\Delta\gamma)^2-16J^2}$, whose real or imaginary character determines whether the reliability dynamics is overdamped or underdamped.

%===============================================================================
% Subsection 5.4:Underdamped regime: oscillatory relaxation
%===============================================================================
\subsection{Underdamped regime: oscillatory relaxation}
\label{subsec:underdamped}

We first consider the \emph{underdamped} regime,
\begin{align}
|\Delta\gamma|<4J,
\end{align}
$\Lambda$ becomes purely imaginary. 
Introducing the real frequency
\begin{align}
\Omega=\sqrt{16J^2-(\Delta\gamma)^2}>0,
\qquad
\Lambda=i\Omega,
\label{eq:Omega_def}
\end{align}
the reliability function \eqref{eq:R_ket11_closed} can be written in the manifestly real form
\begin{align}
R(t)
&=e^{-\bar\gamma t}\!\left[
\frac{32J^2}{\Omega^2}
-\frac{2(\Delta\gamma)^2}{\Omega^2}
\cos\!\left(\frac{\Omega t}{2}\right)
\right]-e^{-2\bar\gamma t} \nonumber\\
&=2e^{-\bar\gamma t}\!\left[1+\frac{(\Delta\gamma)^2}{\Omega^2}
\left(1-\cos\!\left(\frac{\Omega t}{2}\right)\right)\right]
-e^{-2\bar\gamma t} \nonumber\\
&=2e^{-\bar\gamma t}\Xi(t)-e^{-2\bar\gamma t},
\label{eq:R_underdamped}
\end{align}
where
\begin{align}
\Xi(t)=1+\frac{(\Delta\gamma)^2}{\Omega^2}
\left(1-\cos\frac{\Omega t}{2}\right).
\end{align}
Differentiating Eq.~\eqref{eq:R_underdamped} gives
\begin{align}
-\dot R(t)
=
2\bar\gamma\,e^{-\bar\gamma t}\Xi(t)
-
\frac{(\Delta\gamma)^2}{\Omega}
e^{-\bar\gamma t}\sin\!\left(\frac{\Omega t}{2}\right)
-
2\bar\gamma\,e^{-2\bar\gamma t},
\label{eq:Rdot_underdamped}
\end{align}
The hazard function then follows as
\begin{align}
h(t)
=
\frac{-\dot R(t)}{R(t)}
=
\bar\gamma
-
\frac{\frac{(\Delta\gamma)^2}{\Omega}
\sin\!\left(\frac{\Omega t}{2}\right)
+\bar\gamma\,e^{-\bar\gamma t}}
{2\Xi(t)-e^{-\bar\gamma t}} .
\label{eq:h_underdamped}
\end{align}

For later analysis it is convenient to introduce the dimensionless variables
\begin{align}
\theta=\frac{\Omega t}{2},
\qquad
\alpha=\frac{(\Delta\gamma)^2}{\Omega^2},
\end{align}
so that
\begin{align}
\Xi(t)=1+\alpha(1-\cos\theta),
\end{align}
and the hazard becomes
\begin{align}
h(t)=
\bar\gamma
-
\frac{\Omega\,\alpha\,\sin\theta+\bar\gamma e^{-\bar\gamma t}}
{2\bigl[1+\alpha(1-\cos\theta)\bigr]-e^{-\bar\gamma t}} .
\label{eq:h_underdamped_compact}
\end{align}

\paragraph*{Trends and properties.}
Equation~\eqref{eq:R_underdamped} shows that the reliability $R(t)$ displays \emph{exponentially damped} oscillations.
The coherent exchange generates the $\cos(\Omega t/2)$ term, while the oscillation contrast is set by $2(\Delta\gamma)^2/\Omega^2$, vanishing continuously as $\Delta\gamma\to0$ (uniform dissipation).
The overall envelope decays as $e^{-\bar\gamma t}$, and the subtraction $-e^{-2\bar\gamma t}$ enforces $R(0)=1$ and becomes negligible at late times.

It is instructive to contrast this behavior with the familiar two-level Rabi problem~\cite{scully1997quantum} under damping: there the \emph{population} oscillations are progressively washed out by dissipation/dephasing, so the oscillation amplitude in the occupation probabilities decays to zero as the system approaches a steady state.
In our case the same is true for $R(t)$ itself, whose oscillation amplitude is exponentially suppressed by the survival factor $e^{-\bar\gamma t}$.
However, the hazard $h(t)=-\dot R(t)/R(t)$ is a \emph{conditional} instantaneous failure rate.
Once the $e^{-2\bar\gamma t}$ term has dropped out, $R(t)$ takes the asymptotic form $R(t)\simeq e^{-\bar\gamma t} f(t)$ with a bounded periodic factor $f(t)$, implying $h(t)=\bar\gamma-\frac{d}{dt}\ln f(t)$.
Therefore $h(t)$ exhibits oscillatory transients superimposed on a baseline of order $\bar\gamma$ and, in the underdamped regime, approaches a \emph{bounded oscillatory function with time-independent amplitude}.
Increasing $J$ enlarges $\Omega$ and reduces the relative oscillation contrast, whereas stronger inhomogeneity (larger $|\Delta\gamma|$) enhances it.
The oscillation period is $T=4\pi/\Omega$.

%===============================================================================
% Subsection 5.5:Overdamped regime
%===============================================================================
\subsection{Overdamped regime: multirate relaxation}
\label{subsec:overdamped}

Next, we consider the \emph{overdamped} regime
\begin{align}
|\Delta\gamma|>4J,
\end{align}
the quantity $\Lambda$ is real and positive.

Starting from Eq.~\eqref{eq:R_ket11_closed} and using
$e^{\pm\Lambda t/2}=\cosh(\Lambda t/2)\pm\sinh(\Lambda t/2)$,
the reliability can be written in the manifestly real form
\begin{align}
R(t)
&=
e^{-\bar\gamma t}\!\left[
\frac{2(\Delta\gamma)^2}{\Lambda^2}\cosh\!\left(\frac{\Lambda t}{2}\right)
-\frac{32J^2}{\Lambda^2}
\right]
-e^{-2\bar\gamma t}.
\label{eq:R_overdamped_raw}
\end{align}
Using $\Lambda^2=(\Delta\gamma)^2-16J^2$, this expression simplifies to
\begin{align}
R(t)=2e^{-\bar\gamma t}\,\Theta(t)-e^{-2\bar\gamma t},
\label{eq:R_overdamped}
\end{align}
where
\begin{align}
\Theta(t)=1+\frac{(\Delta\gamma)^2}{\Lambda^2}
\left[\cosh\!\left(\frac{\Lambda t}{2}\right)-1\right].
\label{eq:R_overdamped_compact}
\end{align}
Differentiating Eq.~\eqref{eq:R_overdamped} yields
\begin{align}
-\dot{R}(t)
&=
2\bar\gamma e^{-\bar\gamma t}\Theta(t)
-\frac{(\Delta\gamma)^2}{\Lambda}
e^{-\bar\gamma t}\sinh\!\left(\frac{\Lambda t}{2}\right)
-2\bar\gamma e^{-2\bar\gamma t}.
\label{eq:Rdot_overdamped}
\end{align}
The hazard therefore takes the closed form
\begin{align}
h(t)
=
\bar\gamma-
\frac{
\frac{(\Delta\gamma)^2}{\Lambda}\sinh\!\left(\frac{\Lambda t}{2}\right)
+\bar\gamma e^{-\bar\gamma t}
}{
2\Theta(t)-e^{-\bar\gamma t}
}.
\label{eq:h_overdamped}
\end{align}

Introducing the dimensionless variables
\begin{align}
\theta=\frac{\Lambda t}{2},\qquad
\alpha=\frac{(\Delta\gamma)^2}{\Lambda^2},
\end{align}
one may rewrite Eq.~\eqref{eq:h_overdamped} compactly as
\begin{align}
h(t)=
\bar\gamma-
\frac{\Lambda\alpha\sinh\theta+\bar\gamma e^{-\bar\gamma t}}
{2\left[1+\alpha(\cosh\theta-1)\right]-e^{-\bar\gamma t}}.
\label{eq:h_overdamped_compact}
\end{align}

At long times the slowest decay mode in Eq.~\eqref{eq:R_ket11_closed} is
$e^{-(\bar\gamma-\Lambda/2)t}$, implying
\begin{align}
R(t)\sim
\frac{(\Delta\gamma)^2}{\Lambda^2}
\,e^{-(\bar\gamma-\frac{\Lambda}{2})t},
\qquad (t\to\infty).
\end{align}
Consequently, the hazard approaches the constant
\begin{align}
h(t)\longrightarrow
\bar\gamma-\frac{\Lambda}{2}.
\label{eq:h_overdamped_asymp}
\end{align}

\begin{figure*}[t]
\centering

\begin{subfigure}{0.32\textwidth}
\centering
\includegraphics[width=\linewidth]{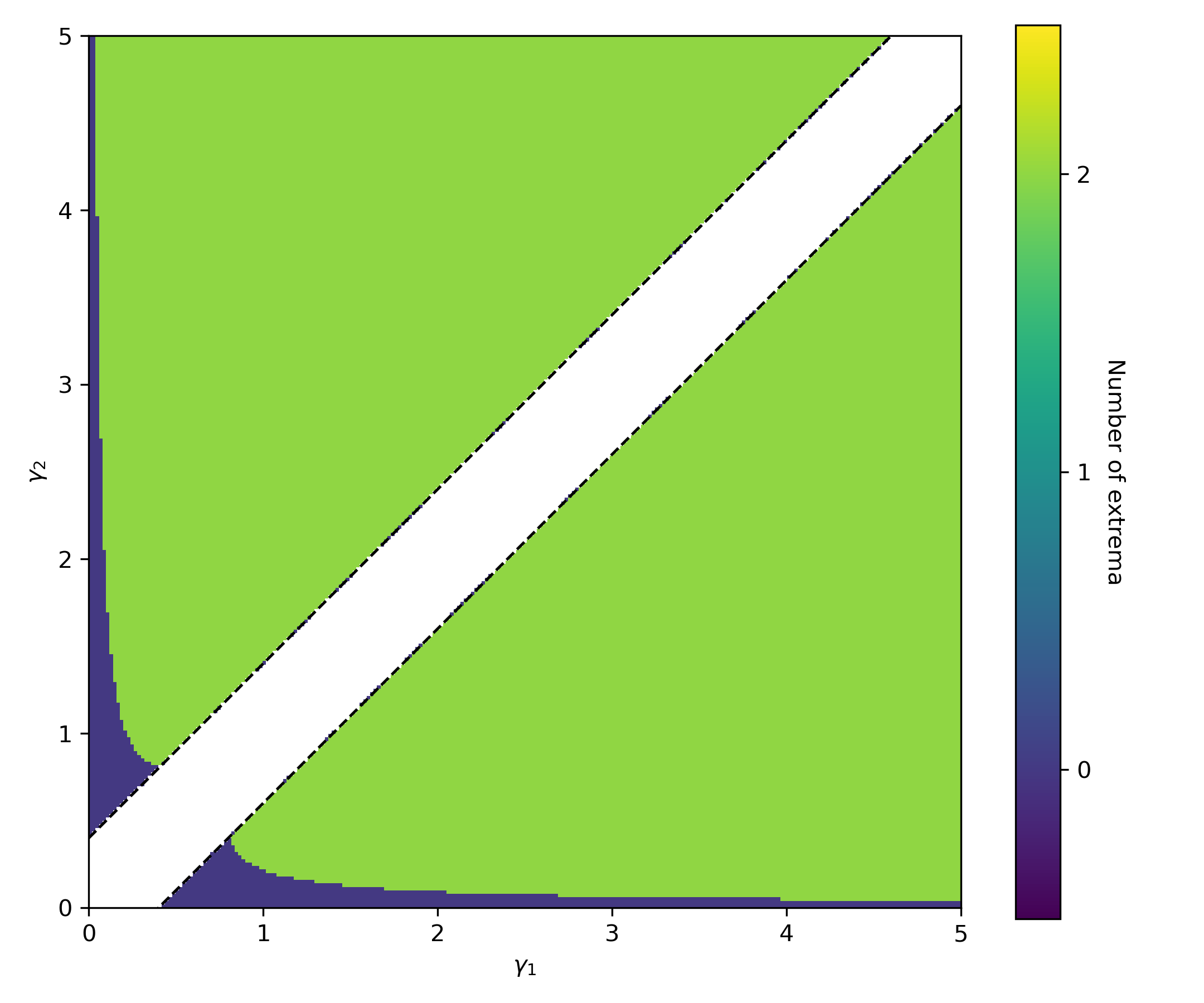}
\caption{$J=0.1$}
\end{subfigure}\hfill
\begin{subfigure}{0.32\textwidth}
\centering
\includegraphics[width=\linewidth]{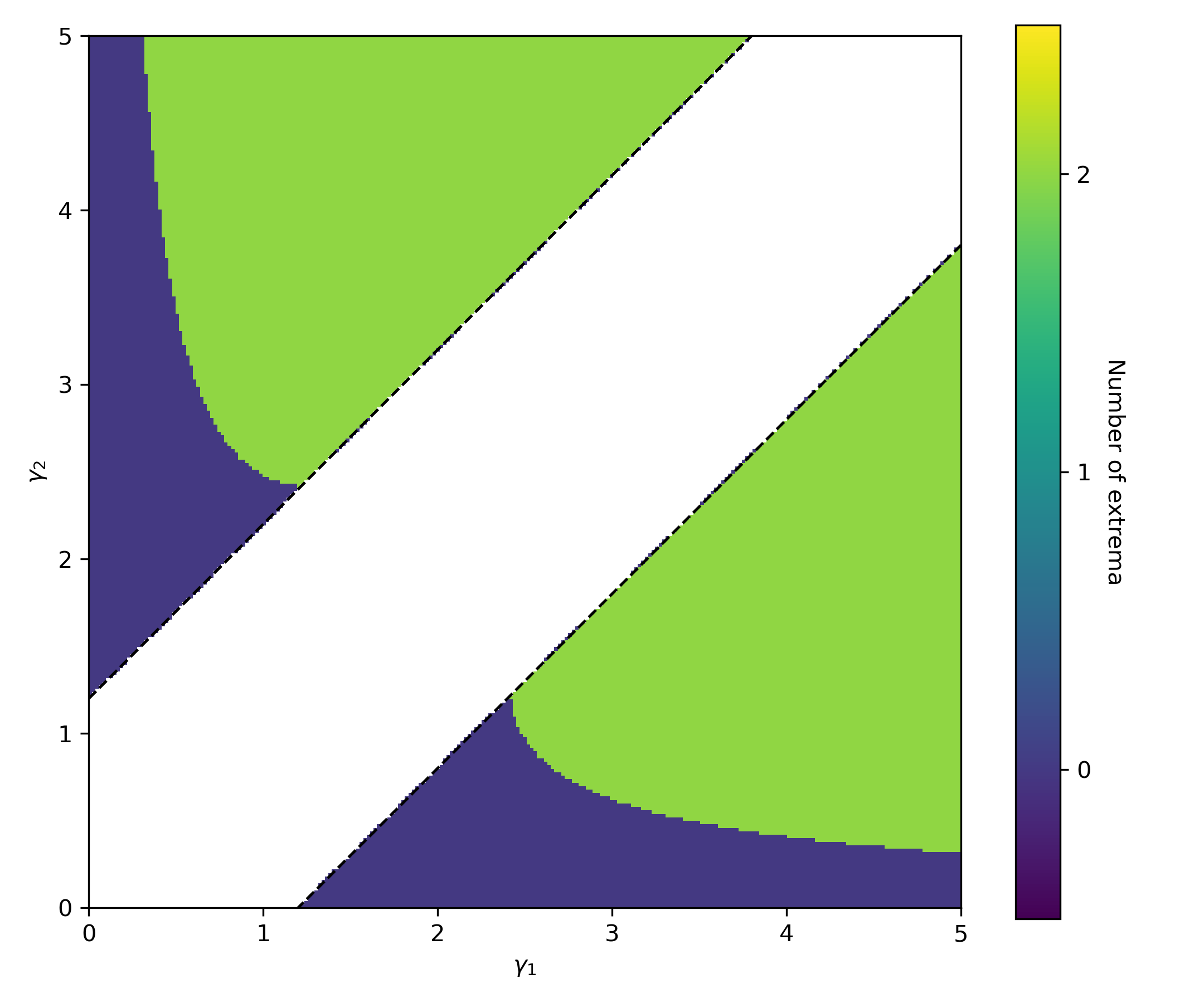}
\caption{$J=0.3$}
\end{subfigure}\hfill
\begin{subfigure}{0.32\textwidth}
\centering
\includegraphics[width=\linewidth]{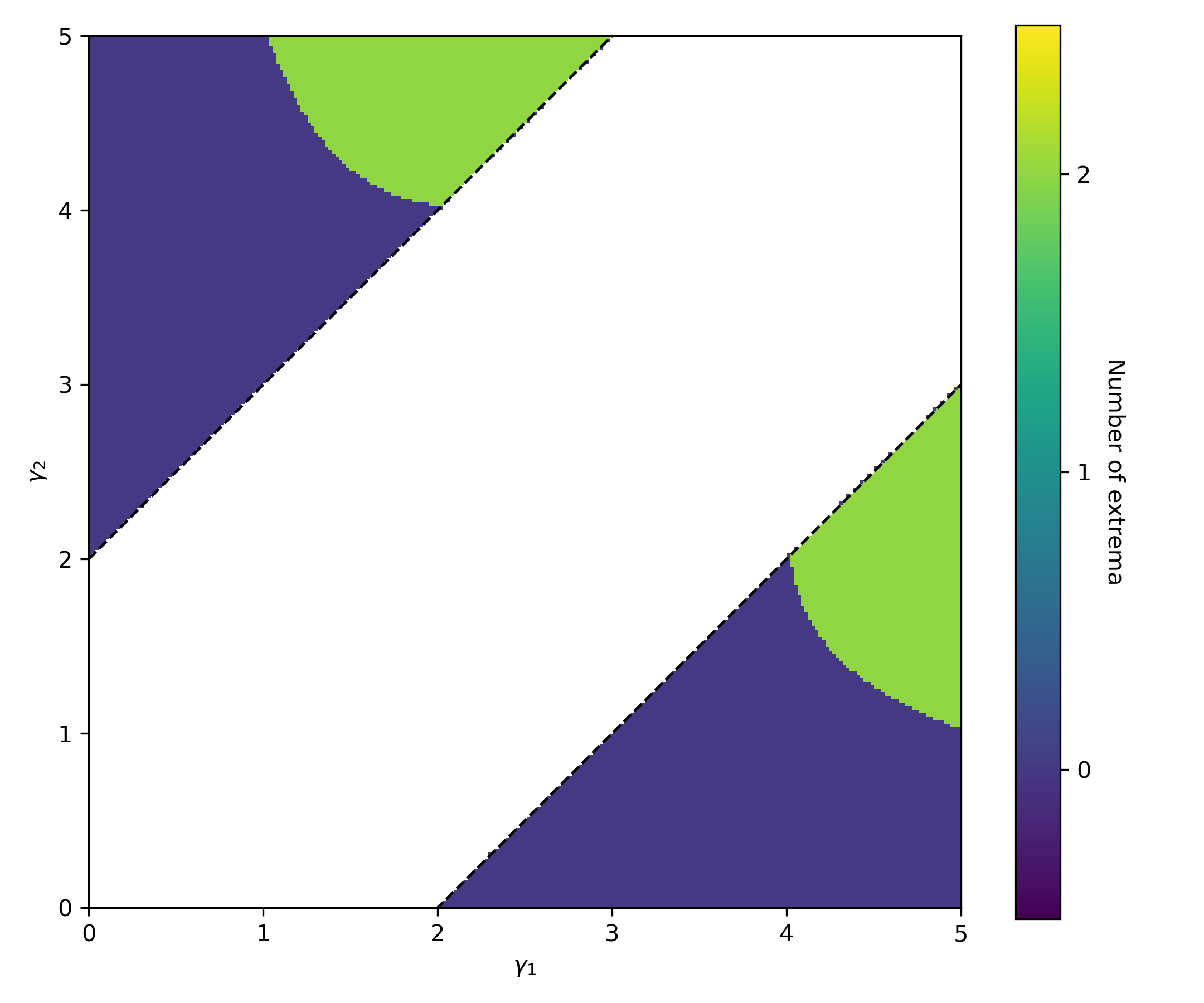}
\caption{$J=0.5$}
\end{subfigure}
\caption{
Extremum-count maps of the overdamped hazard in the $(\gamma_1,\gamma_2)$ plane for (a) $J=0.1$, (b) $J=0.3$, and (c) $J=0.5$. Only the overdamped region $|\Delta\gamma|>4J$ is classified; a narrow band around the crossover line $|\Delta\gamma|=4J$ is left blank. Two sectors appear: a $0$-extrema sector, where $h(t)$ increases monotonically toward the plateau $h_\infty=\bar\gamma-\Lambda/2$, and a $2$-extrema sector, where $h(t)$ shows a local maximum and a local minimum before relaxing to the same plateau.
}
\label{fig:phase_overdamped_triptych}
\end{figure*}

\paragraph*{Extremum structure of $h(t)$.}
In the overdamped regime, depending on $(J,\gamma_1,\gamma_2)$, the hazard $h(t)$ may either
(i) increase monotonically toward the asymptotic value~\eqref{eq:h_overdamped_asymp}, or
(ii) display a nonmonotone profile with two extrema-a local maximum followed by a local minimum-before relaxing to the same plateau.
To analyze this structure we impose the extremum condition
\begin{align}
h'(t^\ast)=0,\qquad t^\ast>0.
\end{align}
Using Eq.~\eqref{eq:R_overdamped},
\begin{align}
h(t)
= -\frac{d}{dt}\ln R(t)
= \bar\gamma-\frac{d}{dt}\ln\!\Bigl[2\Theta(t)-e^{-\bar\gamma t}\Bigr],
\end{align}
so that $h'(t^\ast)=0$ is equivalent to
\begin{align}
\frac{d^2}{dt^2}\ln D(t)=0,
\qquad
D(t)\equiv 2\Theta(t)-e^{-\bar\gamma t}.
\label{eq:D_def_revised}
\end{align}
Writing
\begin{align}
D'(t)&=\alpha\Lambda\sinh\theta+\bar\gamma e^{-\bar\gamma t},
\label{eq:Dprime_revised}\\
D''(t)&=\frac{\alpha\Lambda^2}{2}\cosh\theta-\bar\gamma^2 e^{-\bar\gamma t},
\label{eq:Ddprime_revised}
\end{align}
the extremum condition becomes
\begin{align}
F(t)\equiv D(t)D''(t)-\bigl[D'(t)\bigr]^2=0.
\label{eq:F_def_revised}
\end{align}
Hence the number of extrema of $h(t)$ is determined by the number of positive roots of $F(t)=0$.
For $\gamma_1,\gamma_2>0$,
\begin{align}
F(0)&=-2\gamma_1\gamma_2<0,\\
F'(0)&=(\gamma_1+\gamma_2)\gamma_1\gamma_2>0,
\label{eq:F0_Fp0}
\end{align}
so $F(t)$ starts negative but initially increases.  
In the long-time limit $t\to\infty$,
\begin{align}
F(t)\sim
-\alpha\Lambda^2\Bigl[(\alpha-1)\cosh\theta-\alpha\Bigr]
<0,
\label{eq:F_asymp_negative}
\end{align}
since $\alpha>1$ and $\theta=\Lambda t/2\to\infty$.  
Thus $F(t)$ is negative both at short and long times.
If $F(t)$ never becomes positive, then $h'(t)>0$ for all $t>0$ and $h(t)$ is monotone.  
If instead $F(t)$ becomes positive at intermediate times, continuity requires that it cross zero at least twice before returning to the negative long-time sector. Consequently $h(t)$ develops an even number of extrema. Appendix~\ref{app:two_extrema_proof} further shows that only the $0$- and $2$-extrema cases are possible. Therefore, whenever the nonmonotone branch occurs, $h(t)$ exhibits exactly two extrema: a local maximum followed by a local minimum.
To visualize this behavior we numerically scan the overdamped region in the $(\gamma_1,\gamma_2)$ plane for different $J$ and construct phase-diagram-style maps of the extremum count, shown in Fig.~\ref{fig:phase_overdamped_triptych}.
As $J$ increases, the $2$-extrema sector is progressively compressed, consistent with stronger exchange partially averaging the dissipation imbalance and suppressing intermediate-time nonmonotonicity.

%===============================================================================
% Figure 1
%===============================================================================

\paragraph*{Trends and properties.}
Equation~\eqref{eq:R_overdamped} shows that $R(t)$ is a multirate sum of decaying exponentials without oscillations. 
In the overdamped regime, both the monotone and nonmonotone hazard profiles originate from the same underlying spectral structure: $h(t)$ is the corresponding time-dependent effective decay rate associated with the redistribution of weight among several overdamped decay channels. 
In all cases, the long-time dynamics are governed by the same slow mode $e^{-(\bar\gamma-\Lambda/2)t}$, so that the hazard relaxes to the common asymptotic value $\bar\gamma-\frac{\Lambda}{2}$.
The inhomogeneity parameter $\alpha=(\Delta\gamma)^2/\Lambda^2$ enhances the relative weight of this slow mode, thereby lowering the asymptotic hazard plateau. 
At short times, the hazard starts from $h(0)=0$ and initially increases with positive slope. 
The qualitative difference between the two overdamped subregions lies in the intermediate-time competition among the decay channels. 
If the transfer of spectral weight from the faster transient modes to the slow mode is sufficiently smooth, then $F(t)=D(t)D''(t)-[D'(t)]^2$ remains nonpositive and $h(t)$ increases monotonically toward $h_\infty$. 
By contrast, if the transient contribution becomes sufficiently strong before the slowest mode takes over, then $F(t)$ turns positive at intermediate times, which forces two zero crossings of the extremum condition~\eqref{eq:F_def_revised}. 
In that case, $h(t)$ develops a local maximum followed by a local minimum before relaxing to the same asymptotic plateau.

%===============================================================================
% Section 6:Numerical Verification
%===============================================================================
\section{Numerical Verification}
\label{sec:numerical_validation}

To verify the analytical predictions derived above, we perform direct numerical simulations of the two-site Lindblad master equation for the initial state $\rho(0)=\ket{11}\!\bra{11}$.
The dynamics is integrated using the QuTiP package~\cite{johansson2012qutip}, and the reliability $R(t)$ and the hazard $h(t)$ are extracted from the density-matrix evolution via Eqs.~\eqref{eq:R_def_here} and \eqref{eq:hazard_general}.
The numerical results are compared with the closed-form predictions obtained in Sec.~\ref{sec:closed_form}, in particular the explicit expressions and regime-dependent structures derived in Secs.~\ref{subsec:init11}--\ref{subsec:overdamped}.

\begin{figure*}[t]
\centering
\begin{subfigure}{0.48\textwidth}
\centering
\includegraphics[width=\linewidth]{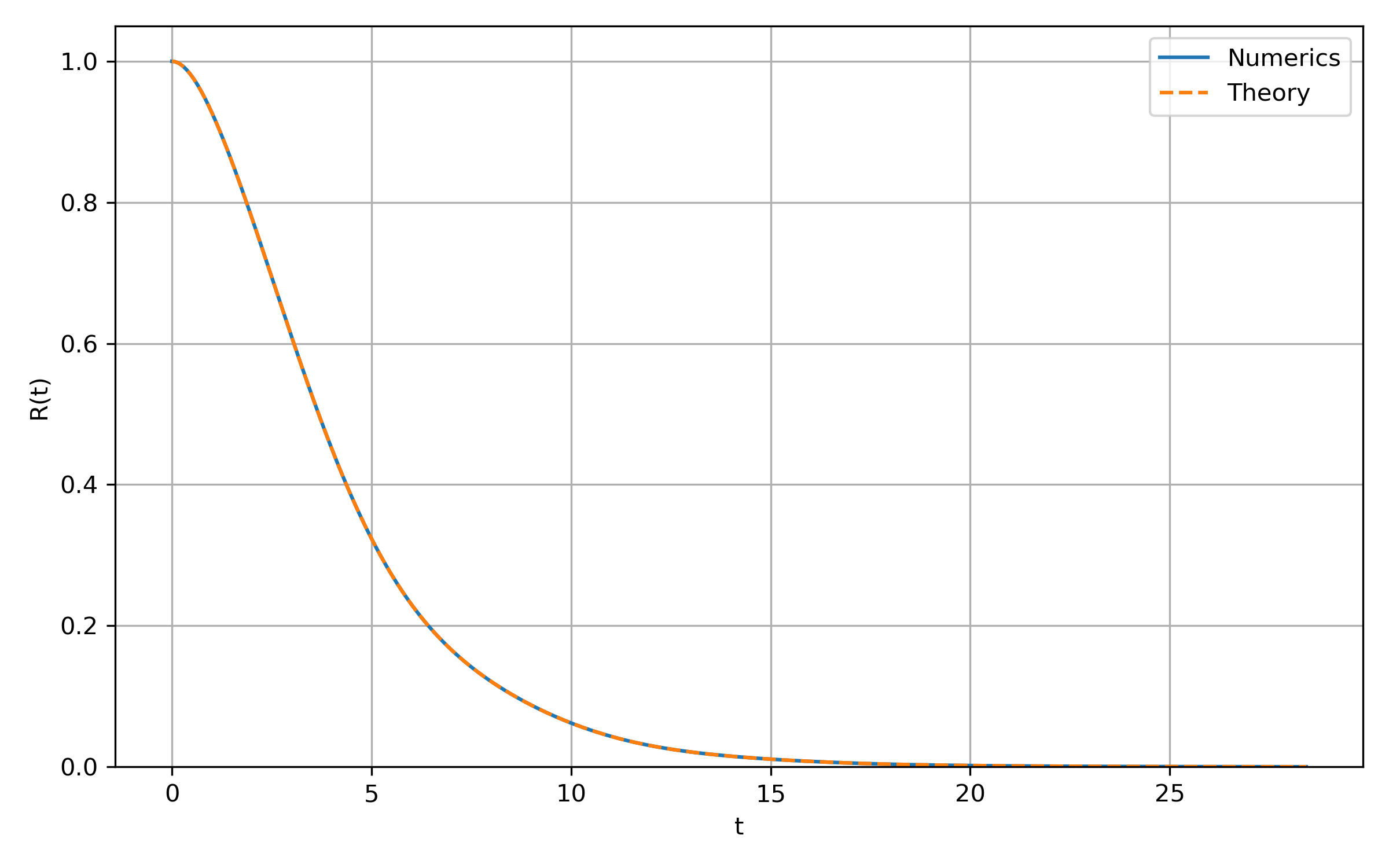}
\caption{$R(t)$ (underdamped).}
\end{subfigure}\hfill
\begin{subfigure}{0.48\textwidth}
\centering
\includegraphics[width=\linewidth]{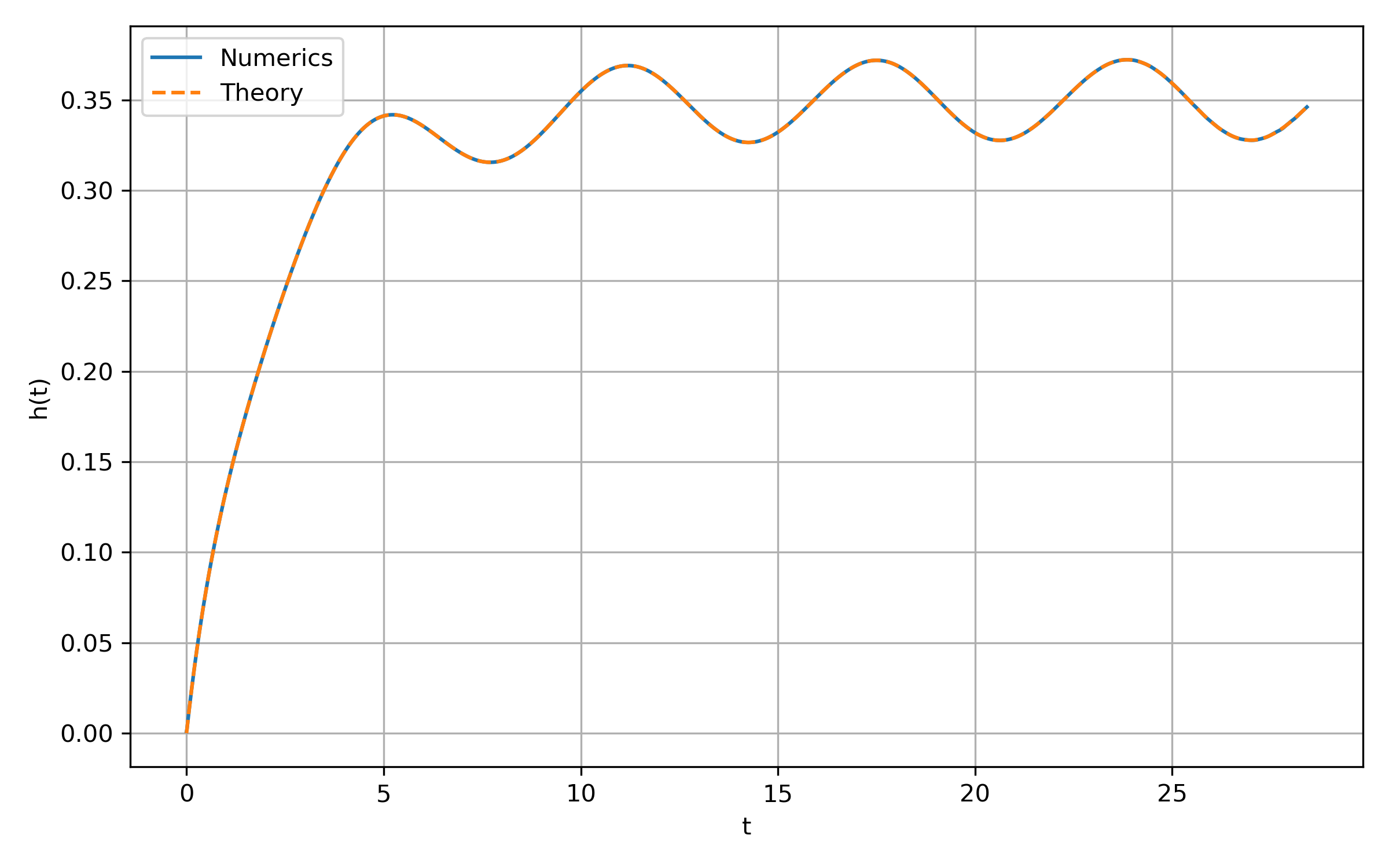}
\caption{$h(t)$ (underdamped).}
\end{subfigure}

\vspace{2mm}

\begin{subfigure}{0.48\textwidth}
\centering
\includegraphics[width=\linewidth]{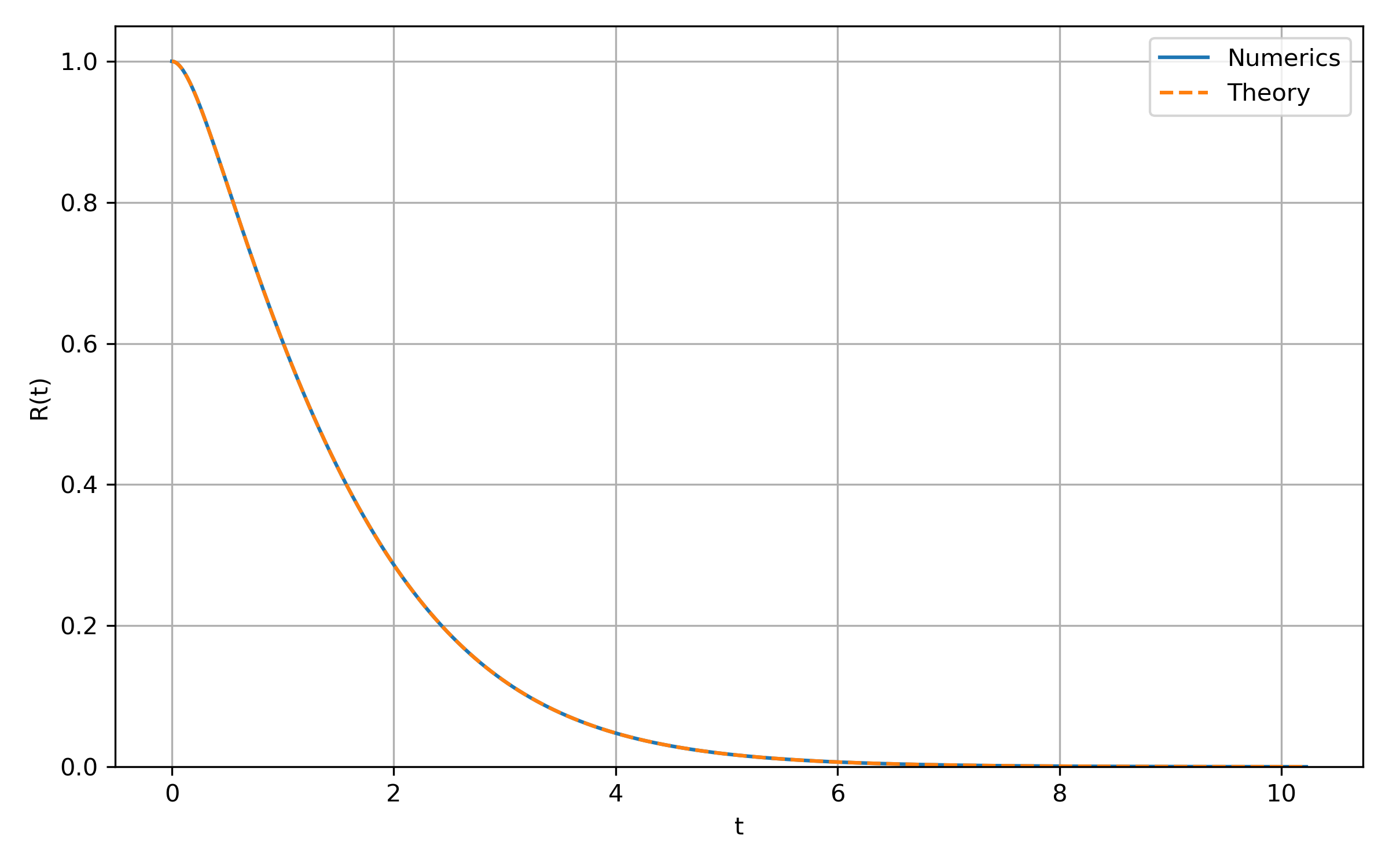}
\caption{$R(t)$ (overdamped, monotone).}
\end{subfigure}\hfill
\begin{subfigure}{0.48\textwidth}
\centering
\includegraphics[width=\linewidth]{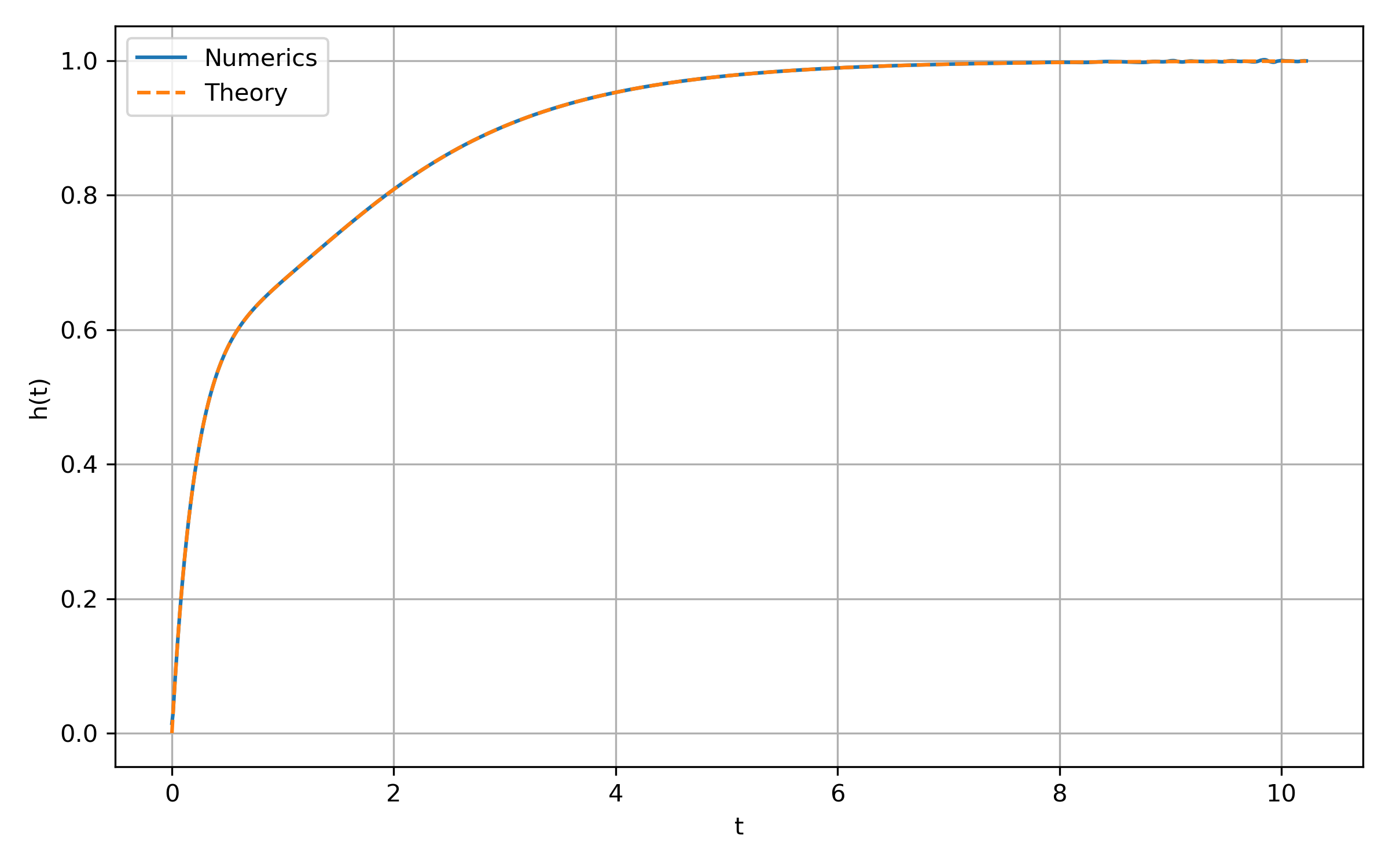}
\caption{$h(t)$ (overdamped, monotone).}
\end{subfigure}

\vspace{2mm}

\begin{subfigure}{0.48\textwidth}
\centering
\includegraphics[width=\linewidth]{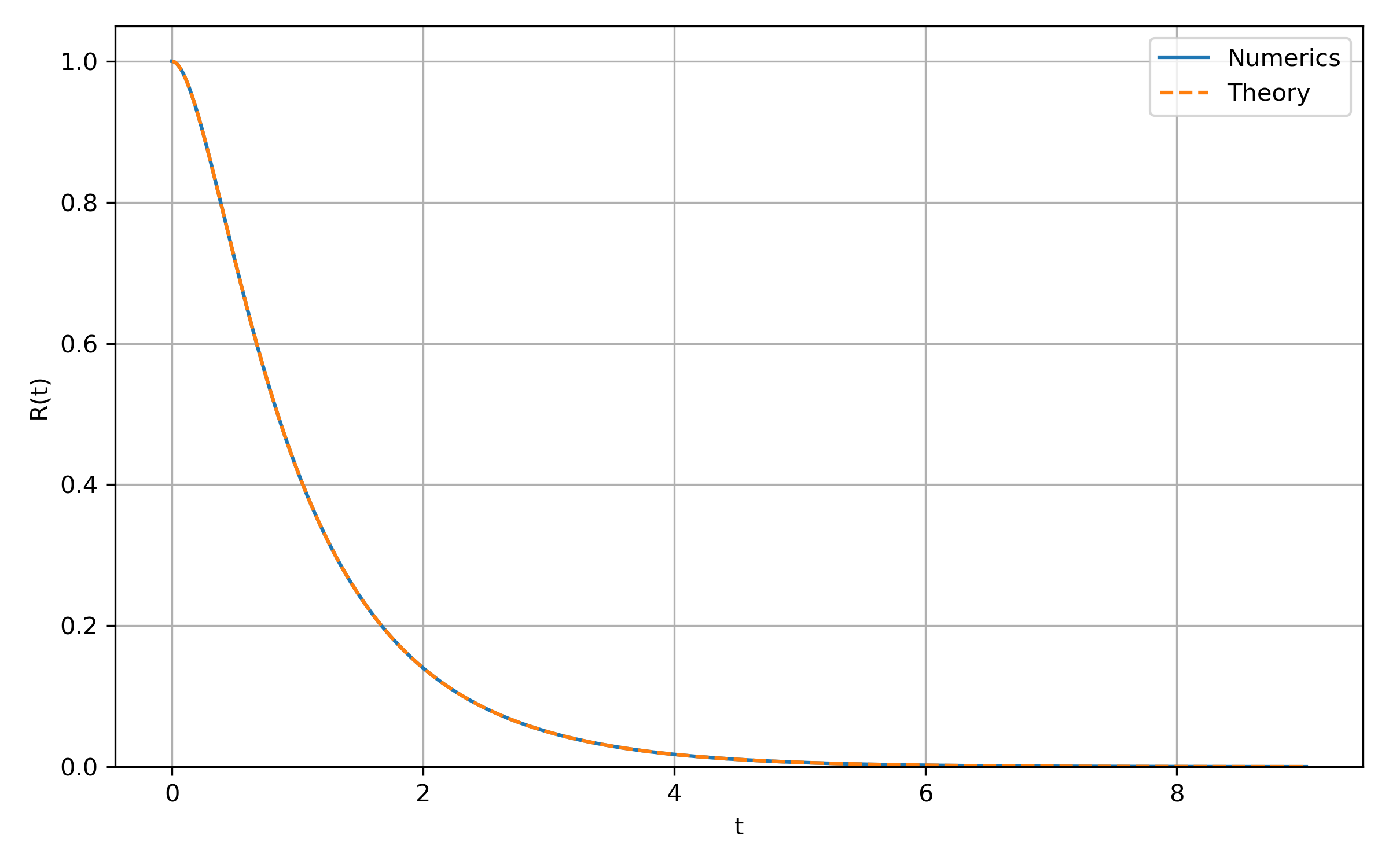}
\caption{$R(t)$ (overdamped, nonmonotone).}
\end{subfigure}\hfill
\begin{subfigure}{0.48\textwidth}
\centering
\includegraphics[width=\linewidth]{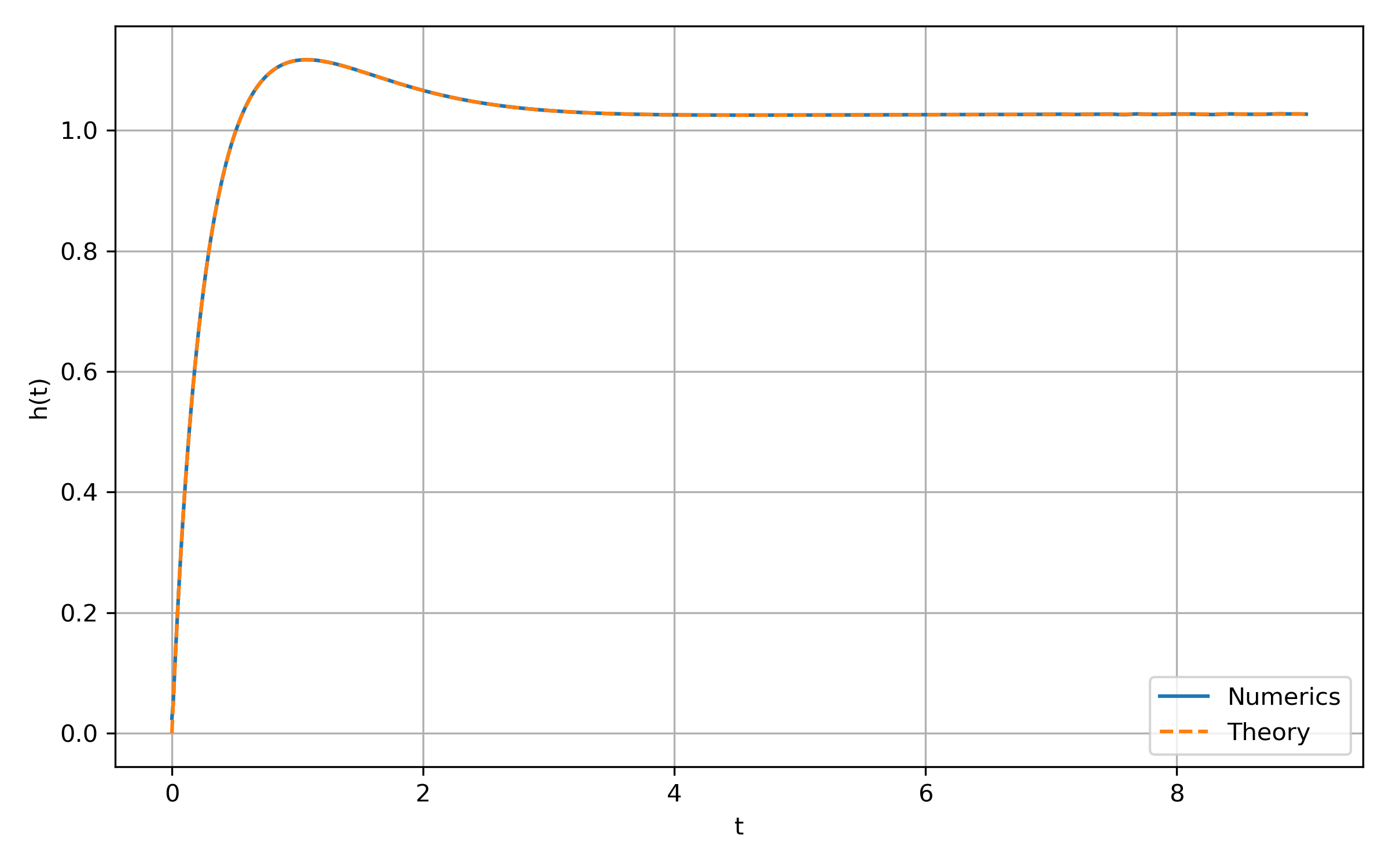}
\caption{$h(t)$ (overdamped, nonmonotone).}
\end{subfigure}

\caption{Numerical verification of the closed-form reliability and hazard for the two-site model initialized in $\ket{11}$. Panels (a,b): underdamped regime, showing oscillatory relaxation. Panels (c,d): overdamped regime with monotone hazard. Panels (e,f): overdamped regime with nonmonotone hazard. Parameters are $J=0.5$, $\gamma_1=0.2$, $\gamma_2=0.5$ for (a,b); $J=0.5$, $\gamma_1=3.0$, $\gamma_2=0.5$ for (c,d); and $J=0.1$, $\gamma_1=2.5$, $\gamma_2=1.0$ for (e,f). Dashed lines denote the analytical results and Solid lines denote Lindblad numerics.}
\label{fig:validation_six}
\end{figure*}

To highlight the dynamical crossover, we consider three representative parameter sets:
(i) an underdamped case with oscillatory $h(t)$;
(ii) a typical overdamped case with monotone $h(t)$; and
(iii) an overdamped case in which $h(t)$ is nonmonotone with two extrema.
These three cases correspond to the six panels of Fig.~\ref{fig:validation_six}, with $R(t)$ and $h(t)$ shown for each parameter set.

In the numerical evaluation of the hazard function, we impose a reliability cutoff to suppress late-time artifacts that arise when $R(t)$ becomes extremely small and the ratio $-\dot R(t)/R(t)$ becomes numerically ill-conditioned.
Specifically, we retain only the time points satisfying
\begin{align}
R(t)>R_{\mathrm{cutoff}}=10^{-4}.
\end{align}
This cutoff removes only the near-zero-reliability region and does not affect the physically relevant part of the dynamics.

%===============================================================================
% Figure 2
%===============================================================================

%===============================================================================
% Section 7: Experimental Assessment via First-Passage Time Statistics
%===============================================================================
\section{Experimental Assessment via First-Passage Time Statistics}
\label{sec:experimental_assessment}

Here, we develop an experimentally accessible protocol for assessing the reliability $R(t)$ and hazard $h(t)$ based on first-passage (first-hitting) time statistics of failure events~\cite{redner2001guide}, thereby avoiding full state tomography.
To this end, we consider a discrete-time monitoring scheme with sampling interval $\Delta t$ and construct practical estimators for $R(t)$ and $h(t)$ directly from the resulting first-passage time data.

%===============================================================================
% Subsection 7.1:Repeated monitoring protocol
%===============================================================================
\subsection{Repeated monitoring protocol}
\label{subsec:fpt_protocol}

We perform $N_s$ independent experimental shots, each initialized in the same state (e.g., $\rho(0)=\ket{11}\!\bra{11}$ for the two-site model), and repeatedly monitor whether the system has entered the failure subspace (for $N=2$, the failure state is $\ket{00}$).
Measurements are applied stroboscopically with a fixed sampling interval $\Delta t$ at times
\begin{align}
t_k \equiv k\,\Delta t,\qquad k=0,1,2,\dots .
\label{eq:tk_def}
\end{align}
For each shot, the experiment terminates once failure is first detected.

For shot $i=1,\dots,N_s$, this procedure yields a measured first-passage time
\begin{align}
T_i^{\rm meas}\in[t_k,t_{k+1}),
\label{eq:Ti_meas_bin}
\end{align}
which corresponds to a discretized (binned) version of the underlying continuous first-passage time.
The temporal resolution therefore improves as $\Delta t\to 0$.

Consequently, the empirical reliability curve reconstructed from $\{T_i^{\rm meas}\}$ is step-like, and its deviation from the true reliability $R(t)$ increases with the sampling interval $\Delta t$.

%===============================================================================
% Subsection 7.2:Discrete-time Reliability estimator
%===============================================================================
\subsection{Discrete-time reliability estimator}
\label{subsec:fpt_R_est}

Given the measured first-passage times $\{T_i^{\rm meas}\}$, a natural estimator of the reliability at the sampling times $\{t_k\}$ is the empirical survival fraction,
\begin{align}
\widehat R(t_k)
=\frac{1}{N_s}\sum_{i=1}^{N_s}\mathbf{1}\!\left(T_i^{\rm meas}>t_k\right),
\label{eq:Rhat_indicator}
\end{align}
where $\mathbf{1}(\cdot)$ denotes the indicator function.
Equivalently, if $n_k$ is the number of failures recorded in the interval $(t_k,t_{k+1}]$, then the number of shots still at risk at time $t_k$ is
\begin{align}
n_{\rm risk}(t_k)=N_s-\sum_{j=0}^{k-1} n_j,
\label{eq:nrisk_def}
\end{align}
so that
\begin{align}
\widehat R(t_k)=\frac{n_{\rm risk}(t_k)}{N_s}.
\label{eq:Rhat_nrisk}
\end{align}

\begin{figure*}[t]
\centering
\begin{subfigure}{0.48\textwidth}
\centering
\includegraphics[width=\linewidth]{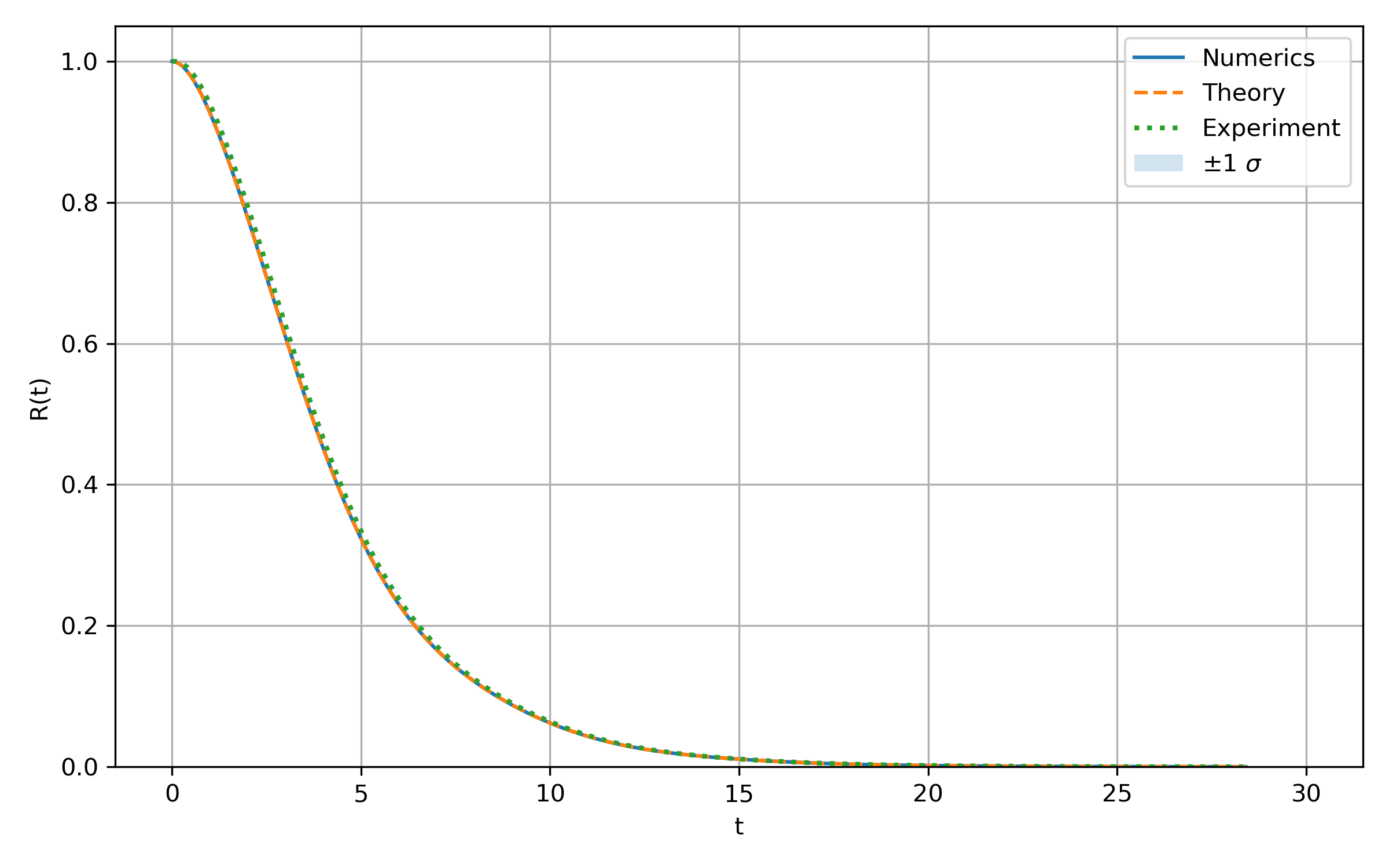}
\caption{Reconstructed $\widehat R(t_k)$ for the underdamped case $J=0.5$, $\gamma_1=0.2$, and $\gamma_2=0.5$ ($|\Delta\gamma|<4J$).}
\label{fig:fpt_R_under}
\end{subfigure}\hfill
\begin{subfigure}{0.48\textwidth}
\centering
\includegraphics[width=\linewidth]{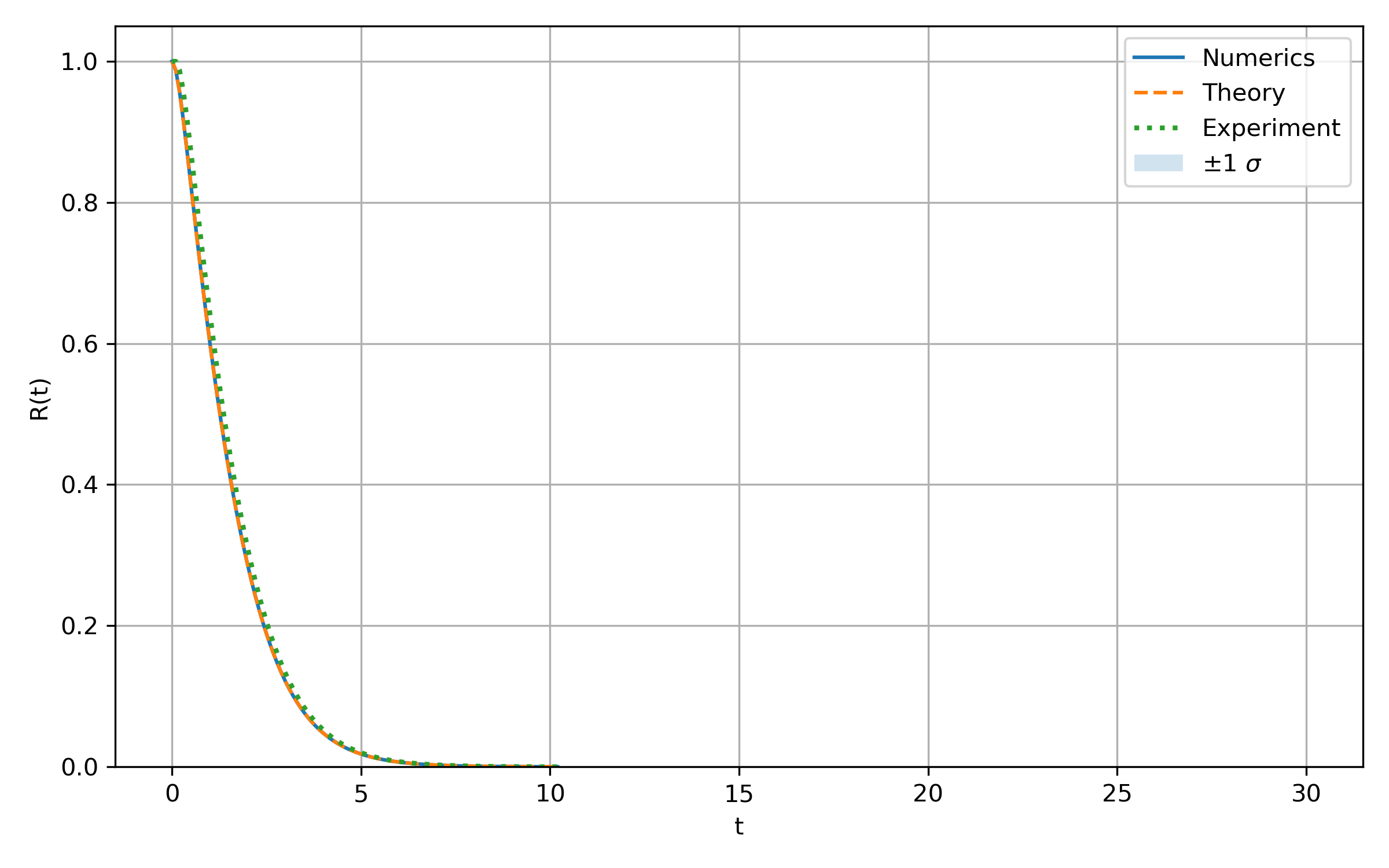}
\caption{Reconstructed $\widehat R(t_k)$ for the overdamped case $J=0.5$, $\gamma_1=3.0$, and $\gamma_2=0.5$ ($|\Delta\gamma|>4J$).}
\label{fig:fpt_R_over}
\end{subfigure}
\caption{Experimental assessment of the reliability $\widehat R(t_k)$ in the underdamped and overdamped regimes, obtained from $N_s=10^6$ independent experimental shots.}
\label{fig:fpt_R}
\end{figure*}

To illustrate the reconstruction of reliability from first-passage data, we first present the estimator $\widehat R(t_k)$ for representative underdamped and overdamped parameter sets.
As shown in Fig.~\ref{fig:fpt_R}, the reconstructed reliability curves are in good agreement with the theoretical predictions in both the underdamped and overdamped regimes, confirming that the first-passage-time protocol faithfully reproduces the reliability dynamics.

%===============================================================================
% Figure 3
%===============================================================================

%===============================================================================
% Subsection 7.3:Discrete-time hazard estimator and its variance
%===============================================================================
\subsection{Discrete-time hazard estimator and its variance}
\label{subsec:fpt_h_est}

On the sampling grid $\{t_k\}$, the hazard can be estimated from its defining interpretation as the failure probability per unit time conditioned on survival up to $t_k$.
This leads to the discrete-time estimator
\begin{align}
\widehat h(t_k)
=\frac{n_k}{\Delta t\,n_{\rm risk}(t_k)} 
=\frac{n_k}{\Delta t\,N_s\,\widehat R(t_k)} ,
\label{eq:hhat_discrete}
\end{align}
where $n_k$ is the number of failures recorded in the interval $(t_k,t_{k+1}]$.
In the limit $\Delta t\to 0$, the estimator $\widehat h(t_k)$ approaches the continuous hazard rate defined by $h(t)=-\dot R(t)/R(t)$.

\begin{figure*}[t]
\centering
\begin{subfigure}{0.48\textwidth}
\centering
\includegraphics[width=\linewidth]{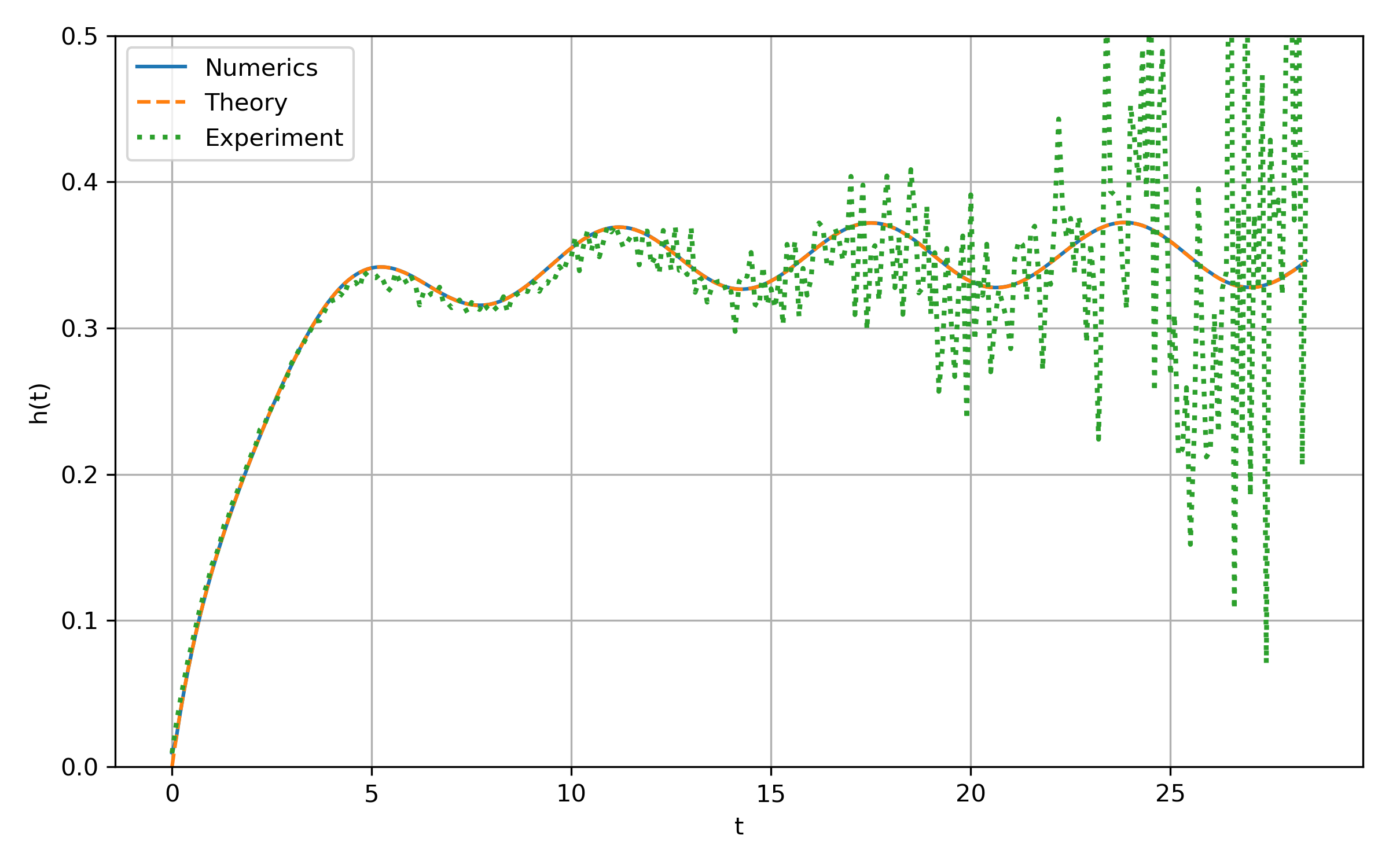}
\caption{Reconstructed $\widehat h(t_k)$ for the underdamped case.}
\label{fig:fpt_h_under}
\end{subfigure}\hfill
\begin{subfigure}{0.48\textwidth}
\centering
\includegraphics[width=\linewidth]{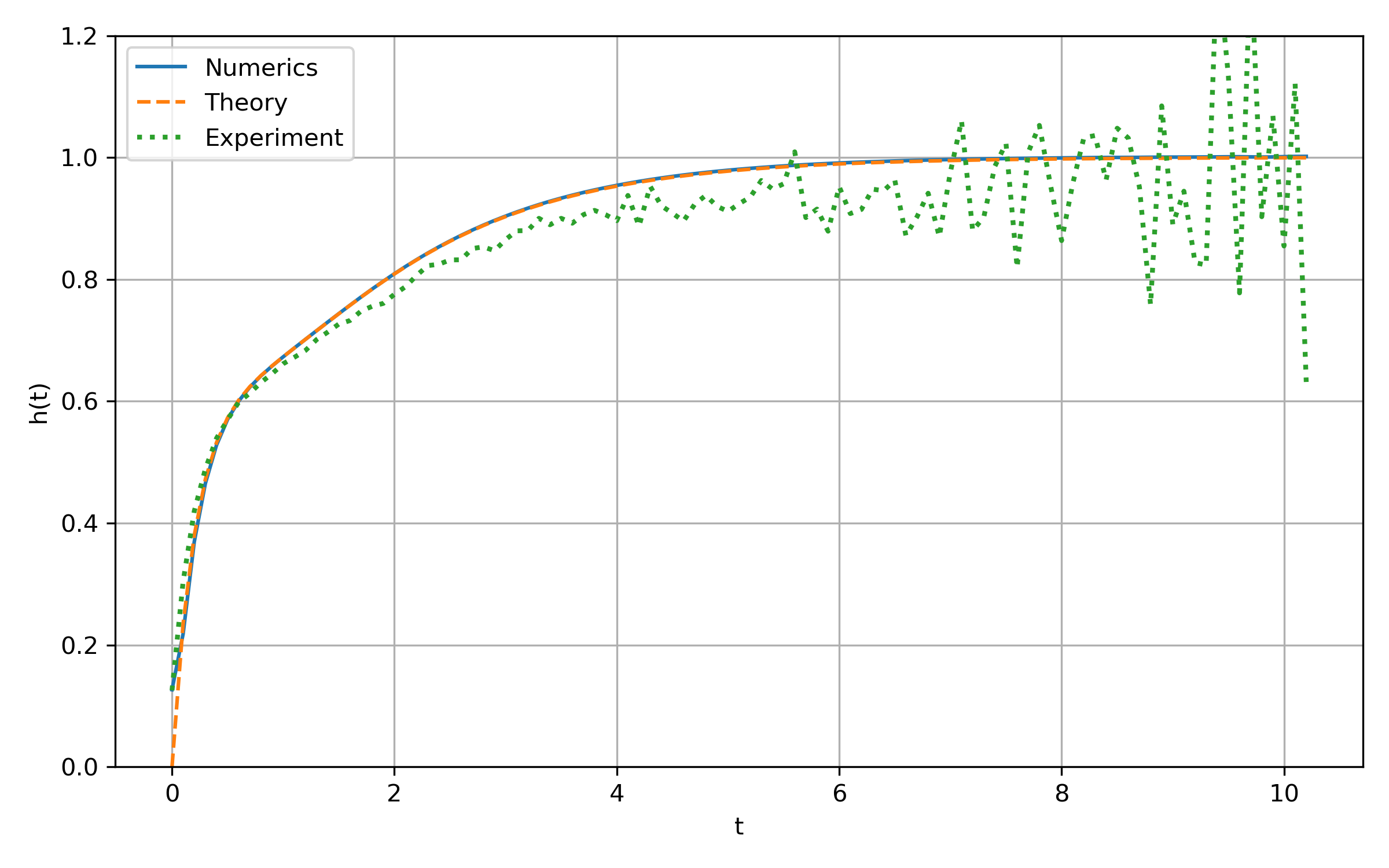}
\caption{Reconstructed $\widehat h(t_k)$ for the overdamped case.}
\label{fig:fpt_h_over}
\end{subfigure}
\caption{Experimental assessment of the hazard $\widehat h(t_k)$ in the underdamped and overdamped regimes, obtained from $N_s=10^6$ independent experimental shots.}
\label{fig:fpt_h}
\end{figure*}

Figure~\ref{fig:fpt_h} shows the reconstructed hazard curves for representative underdamped and overdamped parameter sets.
In the underdamped regime the hazard exhibits a bounded oscillatory modulation, while in the overdamped regime $\widehat h(t_k)$ relaxes toward a constant plateau determined by the slowest decay mode.

%===============================================================================
% Figure 4
%===============================================================================

To quantify sampling fluctuations, note that conditioned on $n_{\rm risk}(t_k)$, the event count $n_k$ can be modeled as a binomial random variable
\begin{align}
n_k \sim {\rm Binomial}\!\bigl(n_{\rm risk}(t_k),\,p_k\bigr),
\label{eq:nk_binomial}
\end{align}
where $p_k$ is the probability that a failure occurs in the interval $(t_k,t_{k+1}]$ given survival up to $t_k$.
For sufficiently small $\Delta t$, the hazard can be regarded as approximately constant within the interval, giving
\begin{align}
p_k \simeq h(t_k)\,\Delta t .
\label{eq:pk_approx}
\end{align}
Using ${\rm Var}(n_k)=n_{\rm risk}(t_k)p_k(1-p_k)$ together with Eq.~\eqref{eq:hhat_discrete}, we obtain
\begin{align}
{\rm Var}\!\bigl[\widehat h(t_k)\bigr]
=\frac{{\rm Var}(n_k)}{\Delta t^{\,2}n_{\rm risk}(t_k)^2}
=\frac{p_k(1-p_k)}{\Delta t^{\,2}\,n_{\rm risk}(t_k)} .
\label{eq:var_hhat_exact}
\end{align}

In the high-resolution limit ($\Delta t\to 0$, so that $p_k\ll1$), this simplifies to
\begin{align}
{\rm Var}\!\bigl[\widehat h(t_k)\bigr]
\simeq \frac{h(t_k)}{\Delta t\,n_{\rm risk}(t_k)}
=\frac{h(t_k)}{\Delta t\,N_s\,\widehat R(t_k)} .
\label{eq:var_hhat_asymp}
\end{align}
Equation~\eqref{eq:var_hhat_asymp} reveals two experimentally relevant scalings:
(i) at fixed $\Delta t$, the variance decreases as $1/N_s$ with the number of experimental shots; and
(ii) at late times the variance increases as $\widehat R(t_k)$ decreases, reflecting the shrinking risk set.

\begin{figure*}[t]
\centering
\begin{subfigure}{0.48\textwidth}
\centering
\includegraphics[width=\linewidth]{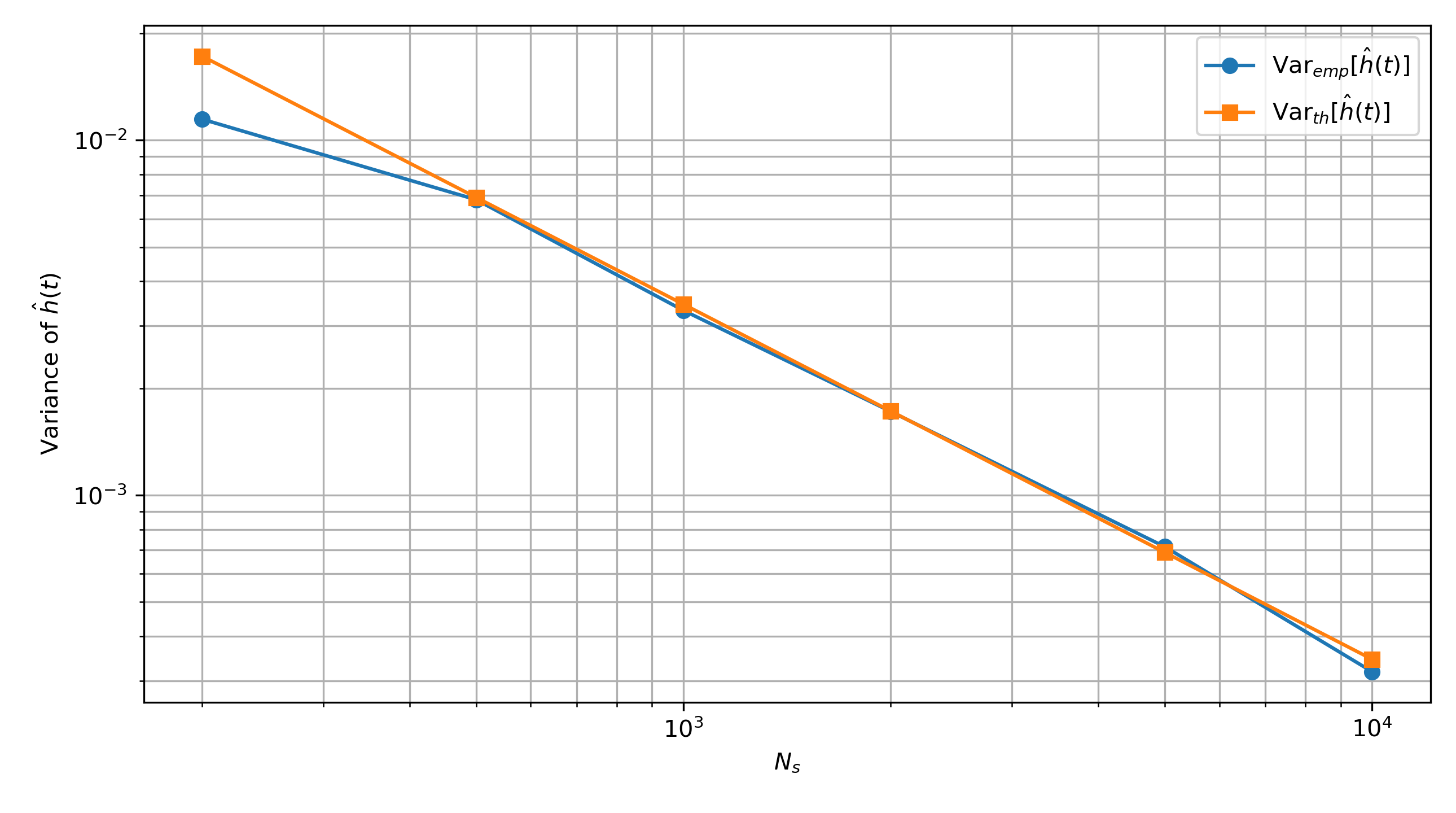}
\caption{$t=2.5$.}
\label{fig:hvar_t25}
\end{subfigure}\hfill
\begin{subfigure}{0.48\textwidth}
\centering
\includegraphics[width=\linewidth]{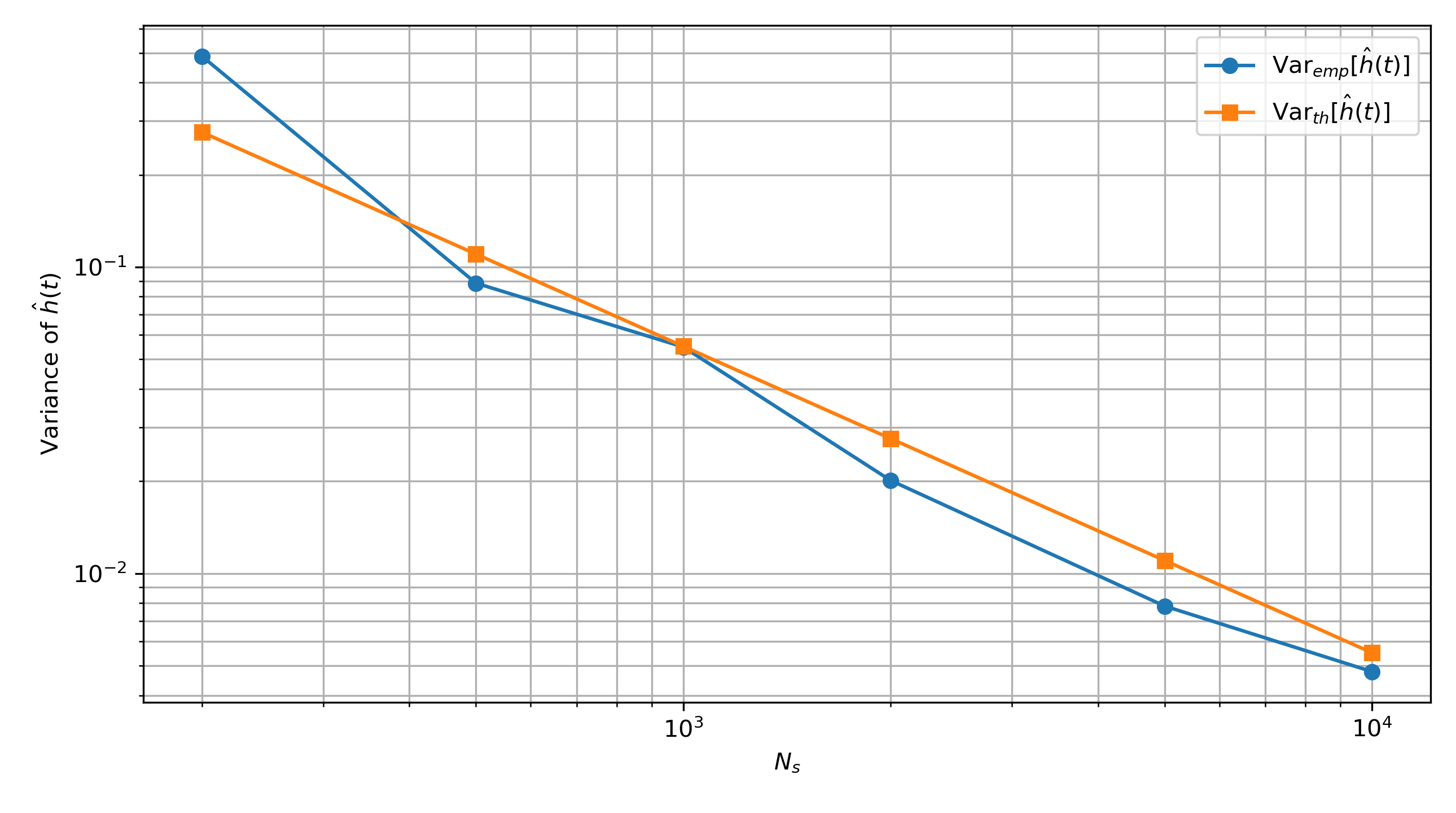}
\caption{$t=10$.}
\label{fig:hvar_t10}
\end{subfigure}

\vspace{2mm}

\begin{subfigure}{0.48\textwidth}
\centering
\includegraphics[width=\linewidth]{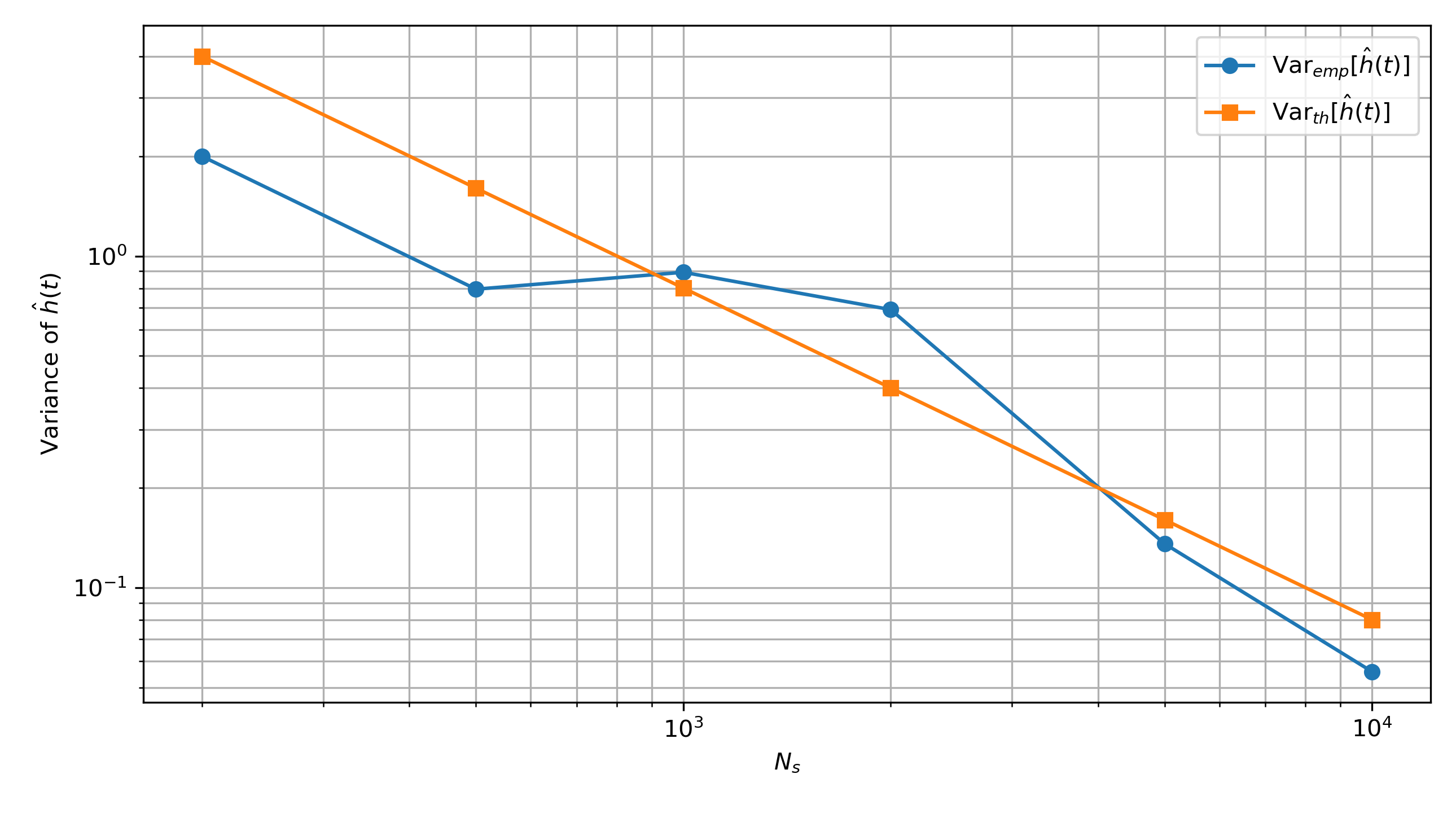}
\caption{$t=17.5$.}
\label{fig:hvar_t175}
\end{subfigure}\hfill
\begin{subfigure}{0.48\textwidth}
\centering
\includegraphics[width=\linewidth]{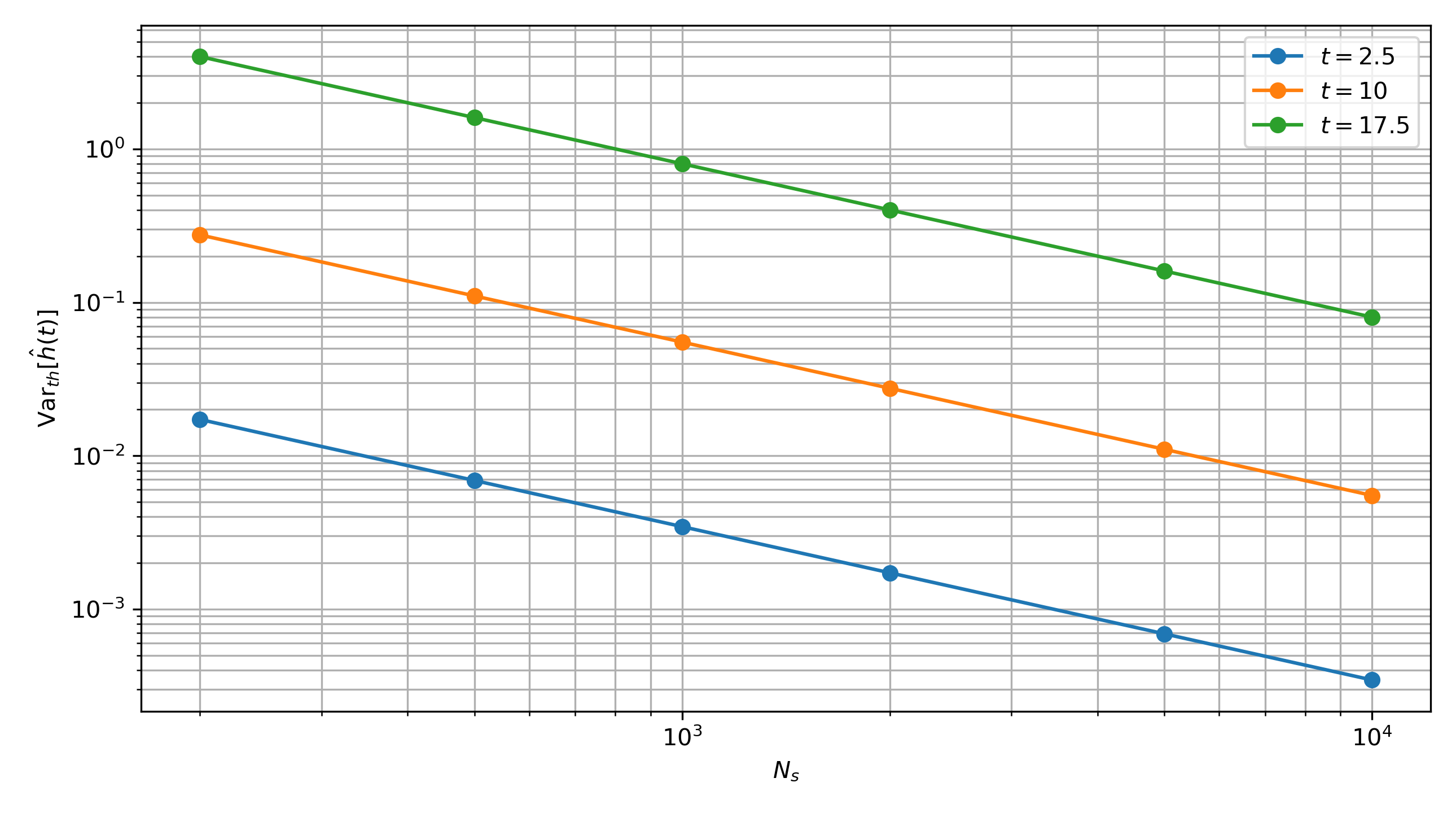}
\caption{Theory-only for $t=2.5,\,10,\,17.5$.}
\label{fig:hvar_theory_only}
\end{subfigure}

\caption{Variance of the discrete-time hazard estimator $\widehat h(t_k)$ versus the number of samples $N_s$ at fixed sampling interval $\Delta t=0.1$. Panels (a)--(c) compare the empirical variance ${\rm Var}_{\rm emp}[\widehat h(t_k)]$, obtained from $50$ independent Monte Carlo repetitions for each $N_s$, with the theoretical prediction ${\rm Var}_{\rm th}[\widehat h(t_k)]$ from Eq.~\eqref{eq:var_hhat_asymp}. Panel (d) shows the theory-only scaling at three representative times, illustrating both the $1/N_s$ decay and the increase of fluctuations at later times due to the shrinking risk set.}
\label{fig:hvar_ntraj}
\end{figure*}

These trends are confirmed numerically in Fig.~\ref{fig:hvar_ntraj}.
Panels (a)--(c) compare the empirical variance ${\rm Var}_{\rm emp}[\widehat h(t_k)]$ with the theoretical prediction ${\rm Var}_{\rm th}[\widehat h(t_k)]$ at three representative times, showing good agreement and an approximately power-law decay with $N_s$ on log--log axes.
Panel (d) summarizes the theory-only curves and highlights the systematic increase of the variance with time due to the decreasing survival probability.

Taking logarithms makes the dominant sample-size scaling explicit,
\begin{align}
\log {\rm Var}\!\bigl[\widehat h(t_k)\bigr]
\simeq -\log N_s + \log\!\left(\frac{h(t_k)}{\Delta t\,\widehat R(t_k)}\right),
\label{eq:logvar_scaling}
\end{align}
so that at fixed $(t_k,\Delta t)$ one expects an approximately unit negative slope in $\log{\rm Var}$ versus $\log N_s$, consistent with the scaling observed in Fig.~\ref{fig:hvar_ntraj}.

%===============================================================================
% Figure 5
%===============================================================================

%===============================================================================
% Subsection 7.4:Practical considerations
%===============================================================================
\subsection{Practical considerations}
\label{subsec:fpt_practical}

The above estimators require only binary failure detection and are therefore much simpler experimentally than full state reconstruction.
Their main limitation is the finite sampling interval: discrete monitoring produces a stepwise $\widehat R(t_k)$ and a piecewise-constant $\widehat h(t_k)$.
Although reducing $\Delta t$ improves temporal resolution, Eq.~\eqref{eq:var_hhat_asymp} shows that it simultaneously increases the estimator variance.
A practical implementation must therefore balance time resolution against statistical fluctuations, and choose $N_s$ large enough that $n_{\rm risk}(t_k)$ remains appreciable throughout the relevant observation window.

%===============================================================================
% Section 8: Conclusion
%===============================================================================
\section{Conclusion}
\label{sec:conclusion}

We have developed a strictly irreversible reliability framework for open quantum spin systems governed by Lindblad amplitude damping, in which the fully decayed ground state acts as an absorbing failure state.
Focusing on the minimal nontrivial benchmark-a two-site spin-$1/2$ chain with coherent exchange and nonuniform local dissipation-we derived closed-form expressions for the reliability $R(t)$ and the hazard $h(t)$ directly from the component equations of motion.

The exact expressions reveal an overdamped-underdamped crossover controlled by the competition between coherent exchange and dissipation inhomogeneity. In the underdamped regime, the reliability exhibits oscillatory relaxation inherited from coherent population exchange. In the overdamped regime, the dynamics is governed by multiple decay modes and the hazard approaches a constant plateau determined by the slowest Liouvillian eigenmode. A detailed analysis of the extremum condition shows that the hazard profile in this regime admits only two possibilities: either it increases monotonically toward its asymptotic value, or it develops exactly two extrema-a local maximum followed by a local minimum-before relaxing to the same plateau. These behaviors, confirmed numerically, establish the two-site model as a transparent and fully solvable benchmark for reliability dynamics in dissipative quantum devices.

Beyond serving as an exactly solvable benchmark, the two-site solution also clarifies mechanisms that are expected to persist in larger and more complex settings. 
In particular, nonuniform dissipation naturally generates multiple decay timescales through the Liouvillian spectrum, suggesting that extended chains with spatially structured decay landscapes may exhibit broad timescale separation, nontrivial finite-size scaling, and possible aging-like behavior in the hazard dynamics. 
Another important direction for future work is to move beyond the effective local-dissipation description adopted here and consider more microscopic environments in which different sites interact with a common or correlated reservoir. 
Such settings can induce bath-mediated cross-dissipative processes and interference between decay channels, potentially leading to qualitatively new reliability dynamics. 
Understanding how these environment-induced correlations reshape the survival probability and hazard structure remains an interesting open problem.

On the experimental side, we proposed an assessment strategy that avoids full state tomography by exploiting first-passage-time statistics under discrete-time monitoring. This protocol provides direct estimators for $\widehat R(t_k)$ and a piecewise-constant $\widehat h(t_k)$, together with an explicit prediction for sampling fluctuations based on a binomial counting model. Such measurements can be implemented on existing platforms including superconducting qubits~\cite{clarke2008superconducting, huang2020superconducting, jiang2025advancements}, trapped ions~\cite{blatt2008entangled, blatt2012quantum, bruzewicz2019trapped}, and Rydberg atom arrays~\cite{browaeys2020many, wu2021concise}.

More broadly, the present work highlights reliability as a quantitative language for assessing the stability of quantum devices. Integrating reliability-based metrics into the modeling and benchmarking of multi-qubit architectures and quantum communication links may provide a useful bridge between microscopic quantum dynamics and engineering-level performance metrics.

\section*{Acknowledgments}

This work was supported by the Science Challenge Project (Grant No.~TZ2025017) and the National Key Research and Development Program of China (Grants No.~2021YFA1402104 and No.~2021YFA0718302).

\appendix

%=====================================================================
% Appendix A: Proof of the 0-or-2 extremum structure of the overdamped hazard function
%=====================================================================
\section{Proof of the \texorpdfstring{$\{0,2\}$}{0,2} extremum structure of the overdamped hazard function}
\label{app:two_extrema_proof}

In the overdamped regime, the nonmonotone branch of the hazard corresponds to exactly two extrema.
A convenient way to expose this structure is to transform the extremum condition into an equation in the variable
\begin{align}
u \equiv e^{-\Lambda t/2}\in(0,1],
\qquad
k \equiv \frac{2\bar\gamma}{\Lambda}.
\label{eq:app_u_def}
\end{align}

Accordingly, the closed-form reliability Eq.~\eqref{eq:R_ket11_closed} can be rewritten as
\begin{align}
R(u)=A\,u^{k-1}+B\,u^k+A\,u^{k+1}-u^{2k}
    =u^{k-1}F(u),
\label{eq:app_Ru}
\end{align}
with
\begin{align}
F(u) \equiv A+Bu+Au^2-u^{k+1},
\label{eq:app_Fu}
\end{align}
where
\begin{align}
A=\frac{(\Delta\gamma)^2}{\Lambda^2}>0,
\qquad
B=-\,\frac{32J^2}{\Lambda^2}<0.
\label{eq:app_AB}
\end{align}

Since $\dot u=-(\Lambda/2)u$, the hazard becomes
\begin{align}
h(t)
=
-\frac{d}{dt}\ln R(t)
=
\frac{\Lambda}{2}
\left[
(k-1)+u\frac{F'(u)}{F(u)}
\right].
\label{eq:app_hu}
\end{align}

Therefore $h'(t^\ast)=0$ for some $t^\ast>0$ is equivalent to
\begin{align}
\frac{dh}{du}(u^\ast)=0,
\qquad
u^\ast\in(0,1).
\end{align}

A direct differentiation yields the scalar extremum equation
\begin{align}
G(u)\equiv uF''(u)F(u)+F'(u)F(u)-u\,[F'(u)]^2=0.
\label{eq:app_G_def}
\end{align}

Thus the number of extrema of $h(t)$ for $t>0$ is exactly the number of
real roots of $G(u)=0$ in the interval $u\in(0,1)$.

\begin{figure}[t]
\centering
\includegraphics[width=0.98\columnwidth]{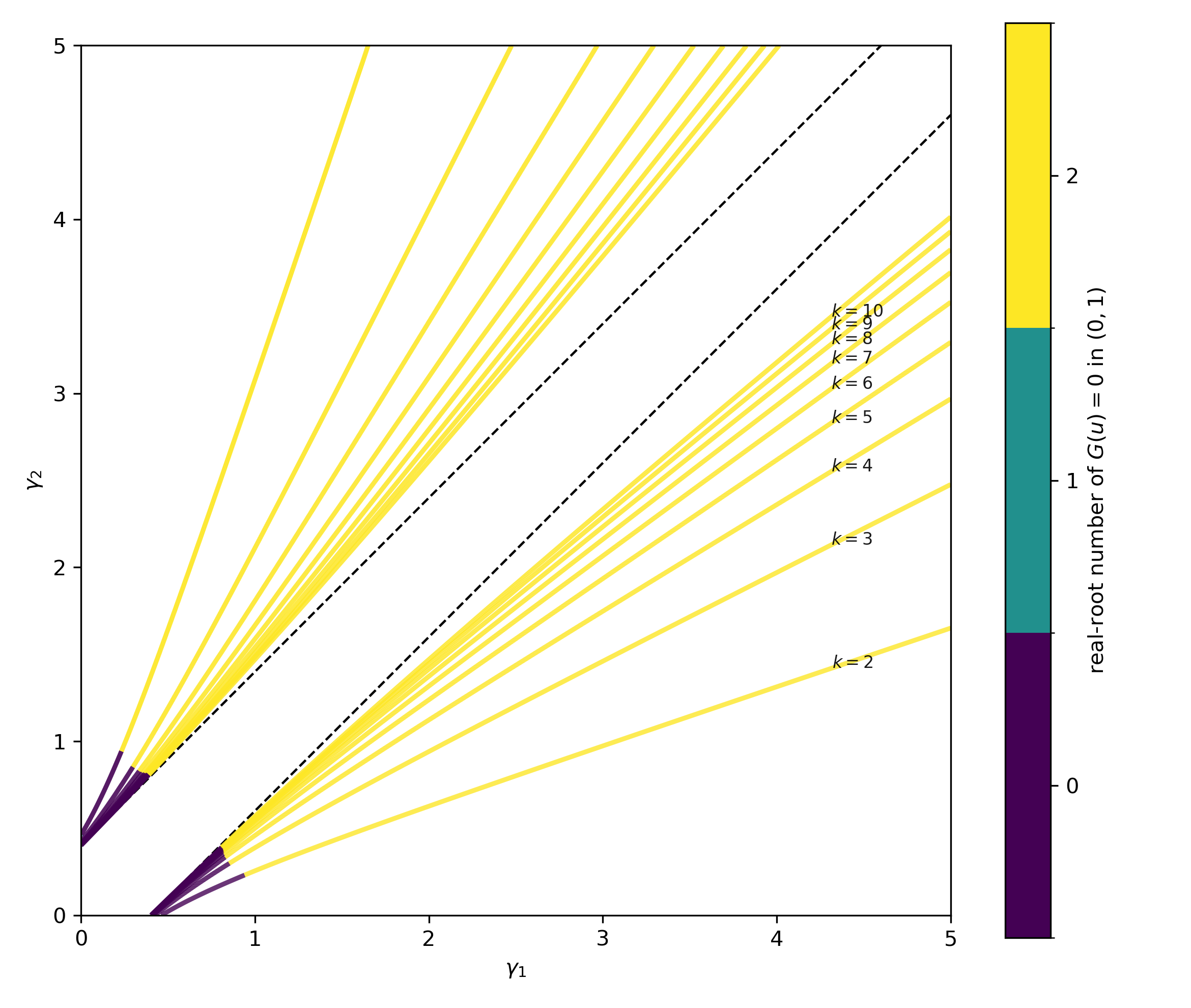}
\caption{
Integer-$n$ curves in the $(\gamma_1,\gamma_2)$ plane for $J=0.1$, colored by the number of real roots of $G(u)=0$ in the interval $(0,1)$.
The dashed lines indicate the overdamped boundaries $|\gamma_1-\gamma_2|=4J$.
Numerically one observes only two sectors: a $0$-root sector and a $2$-root sector.
Since $G(0^+)<0$ and $G(1)<0$, roots can only be created or annihilated in pairs along each integer-$n$ slice, corresponding to the appearance or disappearance of two extrema of the hazard.
}
\label{fig:integer_k_rootcount}
\end{figure}

To gain intuition about the structure of this equation, we first examine the number of real roots of $G(u)=0$ numerically along representative parameter slices.
Figure~\ref{fig:integer_k_rootcount} shows the integer-$k$ curves in the $(\gamma_1,\gamma_2)$ plane together with the number of real roots of $G(u)=0$ in the interval $(0,1)$.
The numerical results indicate that only two possibilities occur: either no real root is present, corresponding to a monotone hazard, or two real roots appear, corresponding to a hazard profile with two extrema.
In particular, no parameter region with a single root is observed.
This empirical observation suggests that extrema of the hazard are created or annihilated in pairs.
In the following analysis we show that this structure follows from the analytic properties of $G(u)$.

Substituting Eq.~\eqref{eq:app_Fu} into Eq.~\eqref{eq:app_G_def} gives
\begin{align}
G(u)
&=AB+4A^2u+ABu^2-A(k+1)^2u^k \nonumber\\
&\quad
-Bk^2u^{k+1}-A(k-1)^2u^{k+2}.
\label{eq:app_G_explicit}
\end{align}

The endpoint values are
\begin{align}
G(0) &= AB < 0,
\label{eq:app_G0}\\
G(1) &= AB+4A^2+AB-A(k+1)^2 \nonumber\\
     &\quad -Bk^2-A(k-1)^2 \nonumber\\
     &= -\frac{8\gamma_1\gamma_2}{\Lambda^2}<0,
\label{eq:app_G1}
\end{align}
where $\gamma_1,\gamma_2>0$ has been used.
Hence, $G(u)$ is negative at both ends of the interval $(0,1)$.
This implies that the number of sign-changing roots of $G(u)=0$ in $(0,1)$ must be even.

For generic parameters, $k$ is noninteger and $G(u)$ is not an ordinary polynomial.
However, when $k=n\in\mathbb{Z}$ with $n\ge 2$, all exponents become integers, so $G(u)$ reduces to an ordinary polynomial of degree $n+2$.
The case $n=1$ is excluded for the present physical parameter range $\gamma_1,\gamma_2>0$, since
\begin{align}
k=\frac{\gamma_1+\gamma_2}{\sqrt{(\gamma_1-\gamma_2)^2-16J^2}}>1
\end{align}
throughout the overdamped regime.

The integer-$n$ slices
\begin{align}
(\gamma_1+\gamma_2)^2
=
n^2\Big[(\gamma_1-\gamma_2)^2-16J^2\Big]
\label{eq:app_integerk_curve}
\end{align}
admit a simple geometric characterization in the $(\gamma_1,\gamma_2)$ plane.
Introducing
\begin{align}
s\equiv \gamma_1+\gamma_2,
\qquad
d\equiv \gamma_1-\gamma_2,
\end{align}
Eq.~\eqref{eq:app_integerk_curve} becomes
\begin{align}
s^2-n^2d^2+16n^2J^2=0,
\label{eq:app_sd_conic}
\end{align}
or equivalently
\begin{align}
\frac{d^2}{(4J)^2}-\frac{s^2}{(4nJ)^2}=1,
\label{eq:app_hyperbola}
\end{align}
which is the standard form of a hyperbola.

The boundaries at which the root count changes are associated with the appearance of a repeated root.
For a polynomial
\begin{align}
P(u)=a_m\prod_{j=1}^{m}(u-u_j),
\end{align}
the discriminant is defined as~\cite{gelfand1994discriminants}
\begin{align}
\mathrm{Disc}(P)
=
a_m^{\,2m-2}\prod_{i<j}(u_i-u_j)^2.
\label{eq:app_disc_def}
\end{align}
Hence $\mathrm{Disc}(P)=0$ if and only if $P$ has a repeated root.
Equivalently one may use the resultant~\cite{cox1997ideals}
\begin{align}
\mathrm{Disc}(P)
=
(-1)^{m(m-1)/2}\,a_m^{-1}\,\mathrm{Res}(P,P').
\label{eq:app_disc_res_relation}
\end{align}

The simplest nontrivial illustration is provided by the slice $k=2$.
Equation~\eqref{eq:app_G_explicit} then becomes
\begin{align}
G(u)=AB+4A^2u+(AB-9A)u^2-4Bu^3-Au^4.
\label{eq:app_G_k2}
\end{align}

Defining
\begin{align}
P(u)\equiv -2G(u),
\end{align}
and parametrizing
\begin{align}
x\equiv -B=\frac{32J^2}{\Lambda^2}>0,
\end{align}
one obtains the quartic
\begin{align}
P(u)
&=
(x+2)u^4-8xu^3+(x^2+11x+18)u^2 \nonumber\\
&\quad
-(2x^2+8x+8)u+(x^2+2x).
\end{align}

Evaluating the resultant $\mathrm{Res}(P,P')$ yields
\begin{align}
\mathrm{Res}(P,P')
&=
-16\,(x-6)\,(x+2)^3
(2x^3-25x^2+68x+124) \nonumber\\
&\quad\times
(x^4-17x^3+91x^2+6x-72).
\end{align}

Within the physical interval $0<x<6$, only the quartic factor produces a real root,
\begin{align}
x^\ast \simeq 0.934756.
\end{align}

This point corresponds to the appearance of a repeated root of $G(u)$ in $(0,1)$ and therefore marks the transition between the $0$-root and $2$-root sectors.

The same mechanism persists for higher integer slices $n\ge2$, for which $G(u)$ remains an ordinary polynomial.
Changes in the number of roots occur only when a repeated root appears inside $(0,1)$.

Finally, the result extends from integer $n$ to noninteger $k$ by continuity.
Since $G(u)$ depends continuously on $k$, the root count can change only through the appearance of a repeated root.
Because the integer slices already contain only the $0$-root and $2$-root sectors, no odd-root sector can emerge upon continuously interpolating to noninteger $k$.

Therefore, throughout the overdamped region, the equation $G(u)=0$ admits only two possibilities in $(0,1)$: either no real root or exactly two real roots.
Since the number of extrema of $h(t)$ equals the number of such roots, the overdamped hazard is either monotone or possesses exactly two extrema-a local maximum followed by a local minimum.

\bibliographystyle{apsrev4-2}
\bibliography{main}

\end{document}